%% file: main.tex
\documentclass[aps,twocolumn, superscriptaddress, footinbib, prb, floatfix]{revtex4-2}

\input{auxiliaries/package_definition}
\input{auxiliaries/tikz_aux}
\input{auxiliaries/additional_declarations}

\def\includeintroduction{1}
\def\includeaszmdef{1}
\def\includeansatz{1}
\def\includecasetwo{1}
\def\includecasethree{1}
\def\includeappendixa{1}
\def\includeappendixb{1}
\def\includecasesamplepic{0}

\usepackage{standalone}

\begin{document}

\title{Variational Minimization of Almost Strong Zero Modes in Parafermionic Wires}

\author{Niklas Tausendpfund}
\affiliation{%
 Forschungszentrum Jülich GmbH, Institute of Quantum Control, Peter Grünberg Institut (PGI-8), 52425 Jülich, Germany
}%
\affiliation{Institute for Theoretical Physics, University of Cologne, D-50937 Cologne, Germany}

\author{Matteo Rizzi}
\affiliation{%
 Forschungszentrum Jülich GmbH, Institute of Quantum Control, Peter Grünberg Institut (PGI-8), 52425 Jülich, Germany
}%
\affiliation{Institute for Theoretical Physics, University of Cologne, D-50937 Cologne, Germany}

\date{\today}

\begin{abstract}

Strongly interacting fermionic chains obeying a $\Z_n$ symmetry can host topological phases with highly degenerate ground-state manifolds. The different ground states are connected by strongly bound fractionalized edge modes called parafermionic zero modes. 
While exactly preserved under time evolution at zero temperature, interactions lead to a destabilization of these modes at finite energy densities, and, given a generic Hamiltonian, obtaining the lifetime of these modes is often impossible as it requires knowledge of the full spectrum of the Hamiltonian or long time evolutions. 
Here, we propose a novel quantum-inspired algorithm that can be used to variationally search for these almost strong zero modes in generic parafermionic chain models.
We demonstrate analytically and numerically that our ansatz faithfully reproduces the exact results and finds strongly bound parafermionic operators in the non-integrable Kitaev-Hubbard chain and chiral Potts model.
Our results further allow us to verify the existence of a true strong zero mode in the symmetrical limit of the chiral Potts model for $\Z_3$-parafermions.

\end{abstract}

\maketitle

\section{\label{sec:intro}Introduction}

\if\includeintroduction1

Coupling an interacting open fermionic wire to a reservoir of Cooper pairs allows for the formation of a novel phase, in which a fermionic degree of freedom fractionalizes between the two ends of the chain.
The two components of this fractionalized fermion are known as Majorana Zero Modes (\textbf{MZM}s)~\cite{Kitaev2001,Ivanov2001,Nayak2008, Alicea2011,Sarma2015}.
This phase is equally described by a two-fold degeneracy in the ground-state manifold~\cite{Stoudenmire2011, Katsura2015}, where two ground states are distinguished only by the occupation of this non-local fermionic mode~\cite{Kitaev2001,Turner2011}.
Consequently, any local measure fails to differentiate between the two ground states, and this phase is known as a symmetry-protected topological phase~\cite{Altland1997, Ryu2010, Turner2011}.
The symmetry protecting this topological degeneracy is the parity of the fermions in the wire, which is characterized by a $\Z_2$ quantum number~\cite{Kitaev2001,Turner2011}.
Moving away from the ground-state manifold, it is questionable how MZMs survive at finite energy densities.
However, it is demonstrated in integrable models that the topological ground-state degeneracy extends throughout the entire many-body spectrum, and the MZM generalizes to a so-called Strong Zero Mode (\textbf{SZM})~\cite{Kitaev2001, Sen13,Fendley2016,Yates2019, Vernier24}.
The existence of such an SZM implies that certain edge operators are preserved under unitary time evolution at arbitrary temperatures~\cite{Else2017, Tausendpfund2025_B}.
Away from the integrable limits, the coupling between the SZM and bulk excitations results in a finite splitting in the spectrum.
Consequently, the lifetime becomes finite but the localized mode survives for orders of magnitude longer timescales than those of any generic bulk excitation ~\cite{Else2017,Parker2019,Kemp2020,Yates2020,Yates2020_B,Yeh2023,Olund2025, Tausendpfund2025}.
These long-lived edge-localized Majorana-like operators are known as Almost Strong Zero Modes (\textbf{ASZMs}) and can be seen as the correct many-body generalizations of MZMs.

By replacing the fermionic parity symmetry with a general $\Z_n$, the concepts of Majorana fermions are generalized to parafermions~\cite{Fendley2012, Alicea2016, Iemini2017_B}.
Similar to the Majorana phase discussed above, parafermionic wires can enter a symmetry-protected topological phase with $n$-fold ground-state degeneracy hosting fractionalized edge excitations~\cite{Fendley2012, Jermyn2014, Calzona2018, Chew2018}.
However, in contrast to MZMs, even the simplest models are non-integrable, and it is not clear how the ground-state degeneracy and edge modes generalize to the many-body spectrum.
In particular, the emergence of SZMs or ASZMs in these wires remains an open question, although strong arguments indicate their existence~\cite{Fendley2012, Jermyn2014}.
The parafermionic wires are dual to spin chains with a local Hilbert space of dimension $n$ by a non-local transformation.
In the dual description, the operators on different sites commute with each other, while possessing complex onsite commutation relations.
We will review the transformation in Section~\ref{sec:parafermion_aszm}.

In this work, we present a novel tensor-network-based algorithm to search for edge-localized parafermionic operators that minimize the commutator with a generic Hamiltonian.
Our algorithm is based on a quantum circuit ansatz that is tailored to operate directly on the correct parafermionic manifold in the dual spin-chain formulation.
This ensures that, during minimization, the ansatz always represents a proper edge-localized parafermionic operator, in contrast to previous work~\cite{Kells2018} which directly approximated the edge operator by a generic Matrix Product Operator (\textbf{MPO}) without fully preserving the desired parafermionic properties.
Using this ansatz, we demonstrate the generic existence of ASZMs in the $n=2$ and $n=3$ parafermionic chain in the topological phase.
We further demonstrate that at the integrable points these ASZMs become true SZMs with an infinite lifetime.

The paper is organized as follows:
In Section~\ref{sec:parafermion_aszm}, we introduce the Hilbert space of $\Z_n$ parafermions and the definition of strong and almost strong zero modes in parafermionic wires.
In Section~\ref{sec:var_ansatz}, we introduce our ansatz to solve the minimization problem for the strong and almost strong zero modes.
Finally, in Sections~\ref{sec:ising_case} and~\ref{sec:chiral_potts_model}, we apply our ansatz to find strong and almost strong zero modes in the case of $n=2$ and $n=3$ parafermions.
We conclude this paper in Section~\ref{sec:concl} by summarizing our findings and outlining future directions.

\fi

\section{\label{sec:parafermion_aszm}Parafermionic Wires and Strong Zero Modes}

\if\includeaszmdef1

In this work, we consider a one-dimensional lattice model of $N$ sites described by two sets of $\Z_n$-parafermionic operators $\chi_j$ and $\psi_j$~\cite{Fradkin1980, Fendley2012}.
These operators have the parafermionic properties for acting on the same lattice site ($\rho_j = \chi_j,\psi_j$)
\begin{equation}\label{eq:parafermions_def_onsite}
    \rho_j^n = \unit\eqcmm\ \rho_j^{(n-1)} = \rho_j^\dag\eqcmm\ \chi_j\psi_j = \omega \psi_j \chi_j\eqcmm\ \omega = e^{2\pi i /n}
\end{equation}
and the commutation relations for $j<k$
\begin{equation}\label{eq:parafermions_def_offsite}
 \rho_j \rho_k = \omega \rho_k\rho_j\eqcmm\ \chi_j \psi_k = \omega \psi_k \chi_j\eqstop
\end{equation}
Based on Eq.~\eqref{eq:parafermions_def_onsite}, it can be shown that the local Hilbert space associated with each lattice site consists of $n$ states~\cite{Fendley2012}.
Thus, the dimension of the many-body Hilbert space $\hilb$ of the $N$-site wire is $d = n^N$.
We assume that the Hamiltonian describing this parafermionic wire is symmetric under a $\Z_n$ symmetry generated by the operator
\begin{equation}\label{eq:paraparity}
    \paraparity = \omega^{-\frac{n-1}{2}N} \prod_{j=1}^N \chi_j \psi_j\eqcmm
\end{equation}
which we call the parafermionic parity operator.
Each of the operators, $\chi_j$ and $\psi_j$, acts individually as a shift operator with respect to this symmetry
\begin{equation}\label{eq:shift_operator_definition}
    \rho_j \paraparity = \omega \, \paraparity \rho_j\eqstop
\end{equation}
Note that for $n=2$ the definition of parafermions reduces to the standard definition of Majorana operators living on a one-dimensional lattice model~\cite{Kitaev2001}.
In this case, the operator $\paraparity$ from Eq.~\eqref{eq:paraparity} becomes the fermionic parity operator up to a global phase.

A parafermionic wire is said to possess a Strong Zero Mode (\textbf{SZM}) $\szm$ if there exists an operator that fulfills the following set of conditions
\begin{enumerate}[label = {C.\arabic*}]
    \item \label{enum:c1} Parafermionic property: $\szm^\dag = \szm^{n-1}$
    \item \label{enum:c2} Normalizability: $\szm^\dag \szm = \unit$
    \item \label{enum:c3} Action as a shift operator: $\szm \paraparity = \omega \, \paraparity \szm$
    \item \label{enum:c4} Localization around position $j = 1$
    \item \label{enum:c5} Commutation with the Hamiltonian for $N\to\infty$: 
    $[H,\szm] = \hat{R}$, $||\hat{R}|| \sim e^{-\lambda N}$, with $\lambda>0$.
\end{enumerate}
Given a Hamiltonian $H$, searching for an SZM is equivalent to solving the linear equation
\begin{equation}\label{eq:szm_linear_equation}
    \liouv\left(\szm\right) \coloneqq [H,\szm] = 0
\end{equation}
under the constraints given by Conditions~\ref{enum:c1}--\ref{enum:c4}.
For example, this was successfully applied in the $n=2$ case to obtain analytical expressions in the integrable limits of the interacting Majorana chain~\cite{Kitaev2001, Fendley2016}.
However, finding a solution in a generic model is difficult and often possible.

Instead of solving the linear equation~\eqref{eq:szm_linear_equation}, identifying an SZM within a generic model can alternatively be formulated as a minimization procedure over the constrained set of operators acting on the parafermionic Hilbert space $\hilb$
\begin{equation}\label{eq:parafermionic_mannifold}
    \paraman \coloneqq \left\lbrace
    \szm: \hilb \to \hilb \  \vert \ \szm\text{ fulfills~\ref{enum:c1}--\ref{enum:c3}}
     \right\rbrace\eqcmm
\end{equation}
referred to as the parafermionic manifold hereafter.
The cost function to be minimized is the commutator functional defined by
\begin{equation}\label{eq:commutator_functional}
     \costfn(\szm) \coloneqq \braket{\liouv(\szm),\liouv(\szm)}\eqcmm\ \szm \in \paraman\eqcmm
\end{equation}
where $\braket{\hat{O},\hat{K}}$ represents the Frobenius scalar product defined by
\begin{equation}\label{eq:frob_inner}
    \braket{\hat{O},\hat{K}} \coloneqq \frac{1}{n^N}\Tr\left[\hat{O}^\dag \hat{K}\right] \eqstop
\end{equation}
Because the Frobenius scalar product is a proper inner product of the vector space formed by the algebra of operators acting on the parafermionic Hilbert space, the cost function in Eq.~\eqref{eq:commutator_functional} is positive semi-definite $\costfn(\szm) \ge 0$.
Therefore, the minimization problem is well defined.
Note that the normalization condition~\ref{enum:c2} is crucial to obtain operators that are defined on the entire Hilbert space and are not simple linear combinations formed inside smaller subspaces, such as the ground-state manifold.
In Section~\ref{sec:var_ansatz}, we introduce an ansatz tailored to solve this minimization problem for modes localized to the first site of the wire (i.e., fulfilling~\ref{enum:c4}).

If the minimum of the commutator functional, Eq.~\eqref{eq:commutator_functional}, saturates to a small but finite number in the thermodynamic limit, i.e., $\min \mathcal{F}(\szm) \xrightarrow{N\to\infty} \mathcal{F}_0>0$, the minimizing operator $\szm\in\paraman$ is called an Almost Strong Zero Mode (\textbf{ASZM}).
In this case, the lifetime $\tau$ of $\szm$ is finite and is inversely proportional to the finite commutator functional~\cite{Else2017} 
\begin{equation}\label{eq:lifetime_aszm}
    \tau \sim \frac{1}{\mathcal{F}_0} \eqstop
\end{equation}
Note that also every operator $\hat{\rho}$ that has a finite overlap with $\szm$ has the same prolonged lifetime.

Before closing this section, we note that the parafermionic Hilbert space on a wire is equivalent to a spin chain of local Hilbert space dimension $n$~\cite{Fendley2012}.
This equivalence is obtained by a non-local unitary transformation 
\begin{equation}\label{eq:jw_traffo_gen}
    \begin{split}
    	\chi_j &= \left(\prod_{k=1}^{j -1} \tau_k \right) \,\sigma_j\\
    	\psi_j &= \omega^{(n-1)/2}\chi_j \tau_j = \omega^{(n-1)/2} \left(\prod_{k=1}^{j -1} \tau_k \right) \,\sigma_j\tau_j \eqcmm
    \end{split}
\end{equation}
mapping the parafermionic operators $\chi_j$ and $\psi_j$ to local spin matrices $\sigma_j$ and $\tau_j$ which commute for different site-indices $j\neq k$.
See Appendix~\ref{app:explicit_representations_spin_mat} for an explicit representation of these spin matrices.
The parafermionic parity operator becomes in the spin picture
\begin{equation}
 \paraparity = \prod_{j=1}^N \tau_j \eqstop   
\end{equation}
Note that for $n=2$, the above expression is nothing other than the well-known Jordan-Wigner transformation, and $\{\sigma,\tau\} \equiv \{\sigma^x, \sigma^z\}$.
%
Since it is simpler to implement spin chains numerically, all the results in this study were obtained in the spin picture.

\fi

\section{\label{sec:var_ansatz}Variational Ansatz}

\if\includeansatz1

To solve the minimization problem in Eq.~\eqref{eq:commutator_functional} over the set defined in Eq.~\eqref{eq:parafermionic_mannifold}, we aim to obtain a tensor network approximation of the operator $\szm$.
To this end, we use the spin language given by Eq.~\eqref{eq:jw_traffo_gen} where all tensors are simply composed of complex numbers and there is no need for graded spaces~\cite{Mortier2025_fermionicTN}.
A naive ansatz would be to set $\szm$ to a general Matrix Product Operator (\textbf{MPO})~\cite{Kells2018}
\begin{equation}
\label{eq:mpo_representation}
    \szm \cong \tikz[baseline=-0.0ex, every node/.style = {anchor = base}]{
        \tikzset{
            wten/.style={
                draw,
                fill=PineGreen!20,
                inner sep = 0,
                minimum width = 0.6cm,
                minimum height = 0.6cm
        }}

        \node[wten] (A1) at (0,0) {$W^1$};
        \node[wten] (A2) at (1.1,0) {$W^2$};
    
        \node (id1) at ($(A1.south) - (0,0.4)$) {};
        \node (id2) at ($(A2.south) - (0,0.4)$) {};

        \node (iu1) at ($(A1.north) + (0,0.4)$) {};
        \node (iu2) at ($(A2.north) + (0,0.4)$) {};

        \node (idr) at ($(A2.east) + (0.3, 0)$) {};
        \node[anchor = center] (dts) at ($(idr) + (0.3, 0)$) {$\dots$};
        \node (idl) at ($(A2.east) + (0.8, 0)$) {};
        \node[wten, anchor = center] (AN) at ($(idl.east) + (0.5, 0)$) {$W^N$};
        \node (idN) at ($(AN.south) - (0,0.4)$) {};
        \node (iuN) at ($(AN.north) + (0,0.4)$) {};

        \draw[thick] (A1) -- (A2);
        \draw[thick] (A2) -- (idr);
        \draw[thick] (idl) -- (AN);

        \draw[thick] (A1) -- (id1) node[below,anchor = center,yshift=-1ex] {$s_1$};
        \draw[thick] (A2) -- (id2) node[below,anchor = center,yshift=-1ex] {$s_2$};
        \draw[thick] (AN) -- (idN) node[below,anchor = center,yshift=-1ex] {$s_N$};

        \draw[thick] (A1) -- (iu1) node[above,anchor = center,yshift=1ex] {$s_1^\prime$};
        \draw[thick] (A2) -- (iu2) node[above,anchor = center,yshift=1ex] {$s_2^\prime$};
        \draw[thick] (AN) -- (iuN) node[above,anchor = center,yshift=1ex] {$s_N^\prime$};
    }\eqstop
\end{equation}
This MPO can be directly deployed to minimize the commutator functional $\costfn(\szm)$ by iteratively minimizing the local tensors until a fixed point is reached, similar to the Density Matrix Renormalization Group (\textbf{DMRG})~\cite{White1992} used to find ground states of a many-body Hamiltonian.
Although this strategy is easy to implement in state-of-the-art simulation platforms and has been used in the past to search for SZMs in Majorana models~\cite{Kells2018}, it has the drawback that the operator $\szm$ is not guaranteed to stay in the parafermionic manifold $\paraman$.
In particular, Condition~\ref{enum:c2} requires the MPO to be a Matrix Product Unitary (\textbf{MPU})~\cite{Cirac2017, Shukla2025}, which is a global condition that cannot be enforced by simple local rules on the tensors $W^j$.
Thus, even if we started with a proper MPU, performing one sweep of the DMRG algorithm would, in the general case, completely destroy the unitary property of our ansatz.

To solve this issue, we propose an alternative tensor network structure that directly incorporates all conditions~\ref{enum:c1}--\ref{enum:c3} in the ansatz, although it requires more sophisticated methods to determine the minimum of the commutator functional $\costfn(\szm)$.
The idea is to dress a seed operator $\szm_0\in\paraman$ by a unitary $\unitary$ that commutes with the parafermionic parity operator $\paraparity$:
\begin{equation}\label{eq:unitary_ansatz}
    \szm(\unitary) = \unitary\szm_0\unitary^\dag\eqcmm \ \left[\unitary, \paraparity\right] = 0\eqstop
\end{equation}
By construction, one has $\szm(\unitary) \in \paraman$, as the unitary $\unitary$ preserves the global properties of the seed operator $\szm_0$.
A suitable seed is then given by an operator that is localized to the first site (\ref{enum:c4}) and that can be written as a MPO of bond dimension one:
\begin{equation}\label{eq:seed_operatator_general}
\szm_0 =  \tikz[baseline=-0.5ex]{
        \tikzset{
            oins/.style={
                circle,
                draw,
                fill=PineGreen!20,
                inner sep = 0,
                minimum width = 0.6cm,
                minimum height = 0.6cm
        }}
        \node[oins] (x) at (0,0) {$\rho$};
        \draw[thick] (x) -- ($(x) + (0.0,0.5)$);
        \draw[thick] (x) -- ($(x) - (0.0,0.5)$);
        \draw[thick] (0.7,0.5) -- (0.7,-0.5);
        \node[] at (1.4,0) {$\dots$};
        \draw[thick] (2.1,0.5) -- (2.1,-0.5);
    }\eqstop
\end{equation}
%

Finally, to arrive at the tensor network representation, we further approximate the full many-body unitary $\unitary$ using a circuit composed of $k$-body unitaries $u_j$ acting on the indices $\lbrace (j-1)k+1, (j-1)k+2, \dots, jk \rbrace$.
The $k$-body unitaries are arranged in a ladder structure~\footnote{It may be more natural to use a brick-wall circuit instead of the ladder structure proposed here. However, because the seed operator in Eq.~\ref{eq:seed_operatator_general} only acts non-trivially on the first site, most of the unitary matrices appearing in the brick wall circuit would cancel, and the effective structure that remains is the circuit proposed here.} as shown here for example for $k = 4$
\begin{equation}\label{eq:ladder_structure}
    \unitary \approx  \tikz[baseline=5.5ex]{
        \tikzset{
            wten/.style={
                draw,
                fill=BurntOrange!20,
                inner sep = 0,
                minimum width = 1.2cm,
                minimum height = 0.6cm
        }}
        \node[wten] (u1) at (0,0) {$u_{1}$};
        \node[wten] (u2) at (1,1) {$u_{2}$};
        \node[wten] (u3) at (2,2) {$u_{3}$};
        
        \draw[thick] ($(u1.south) + (-0.5,-0.4)$) -- ($(u1.south) + (-0.5,0)$);
        \draw[thick] ($(u1.south) + (-0.2,-0.4)$) -- ($(u1.south) + (-0.2,0)$);
        \draw[thick] ($(u1.south) + ( 0.2,-0.4)$) -- ($(u1.south) + ( 0.2,0)$);
        \draw[thick] ($(u1.south) + ( 0.5,-0.4)$) -- ($(u1.south) + ( 0.5,0)$);
        \draw[thick] ($(u1.north) + (-0.5, 2.4)$) -- ($(u1.north) + (-0.5,0)$);
        \draw[thick] ($(u1.north) + (-0.2, 2.4)$) -- ($(u1.north) + (-0.2,0)$);
        \draw[thick] ($(u1.north) + ( 0.2, 2.4)$) -- ($(u1.north) + ( 0.2,0)$);

        \draw[thick] ($(u1.north) + ( 0.5, 0.0)$) -- ($(u2.south) + (-0.5,0)$);
        \draw[thick] ($(u2.south) + (-0.2,-1.4)$) -- ($(u2.south) + (-0.2,0)$);
        \draw[thick] ($(u2.south) + ( 0.2,-1.4)$) -- ($(u2.south) + ( 0.2,0)$);
        \draw[thick] ($(u2.south) + ( 0.5,-1.4)$) -- ($(u2.south) + ( 0.5,0)$);
        \draw[thick] ($(u2.north) + (-0.5, 1.4)$) -- ($(u2.north) + (-0.5,0)$);
        \draw[thick] ($(u2.north) + (-0.2, 1.4)$) -- ($(u2.north) + (-0.2,0)$);
        \draw[thick] ($(u2.north) + ( 0.2, 1.4)$) -- ($(u2.north) + ( 0.2,0)$);

        \draw[thick] ($(u2.north) + ( 0.5, 0.0)$) -- ($(u3.south) + (-0.5,0)$);
        \draw[thick] ($(u3.south) + (-0.2,-2.4)$) -- ($(u3.south) + (-0.2,0)$);
        \draw[thick] ($(u3.south) + ( 0.2,-2.4)$) -- ($(u3.south) + ( 0.2,0)$);
        \draw[thick] ($(u3.south) + ( 0.5,-2.4)$) -- ($(u3.south) + ( 0.5,0)$);
        \draw[thick] ($(u3.north) + (-0.5, 0.4)$) -- ($(u3.north) + (-0.5,0)$);
        \draw[thick] ($(u3.north) + (-0.2, 0.4)$) -- ($(u3.north) + (-0.2,0)$);
        \draw[thick] ($(u3.north) + ( 0.2, 0.4)$) -- ($(u3.north) + ( 0.2,0)$);
        \draw[thick] ($(u3.north) + ( 0.5, 0.4)$) -- ($(u3.north) + ( 0.5,0)$);
        \node[anchor=east] at (3.5,1) {$\dots$};
    }\ \eqstop
\end{equation}
Note that this type of circuit can be directly implemented in a quantum computer by further decomposing the unitaries of block size $k$ into smaller two-body unitaries.
This particular ladder structure is only one possible way of arranging the unitaries.
It is also possible to choose a different form of overlapping between two neighboring unitaries.
Exploring the effect of a different overlap or different circuit-architectures is left for future work.

This ladder structure has the advantage that it preserves locality, in the sense that $\unitary \szm_0 \unitary$ remains exponentially localized to the first site for a localized $\szm_0$ as long as the first unitary has a finite identity component $u_1 = a\unit + \dots$, $ a>0$.
Consequently, for a seed operator as given in Eq.~\eqref{eq:seed_operatator_general}, the ansatz from Eq.~\eqref{eq:unitary_ansatz} still satisfies the localization condition~\ref{enum:c4}.
We want to stress that a similar approach was recently chosen to find the full many-body spectrum of many-body localized Hamiltonians~\cite{Pollmann2016, Wahl2017, Pancotti2018}.

To ensure that the global unitary $\unitary$ commutes with the parafermionic parity, $\left[\unitary, \paraparity\right] = 0$, the local unitaries are chosen to individually commute with the parity $\left[u_j, \paraparity\right] = 0$.
On the level of the tensors, this is explicitly enforced by using symmetric tensors that transform under the action of $\paraparity$ as
\[
    \tikz[baseline=-0.5ex]{
        \tikzset{
            wten/.style={
                draw,
                fill=BurntOrange!20,
                inner sep = 0,
                minimum width = 1.6cm,
                minimum height = 0.6cm},
            oins_loc/.style={
                circle,
                draw,
                fill=Dandelion!20,
                inner sep = 0,
                minimum width = 0.3cm
        }}

        \node[wten] (u1) at (0,0) {$u_j$};
        \node[oins_loc, fill=TealBlue!20] (s1) at ($(u1.north) + (-0.6, 0.4)$) {\small$\tau$};
        \node[oins_loc, fill=TealBlue!20] (s2) at ($(u1.north) + (-0.2, 0.4)$) {\small$\tau$};
        \node[oins_loc, fill=TealBlue!20] (s3) at ($(u1.north) + ( 0.2, 0.4)$) {\small$\tau$};
        \node[oins_loc, fill=TealBlue!20] (s4) at ($(u1.north) + ( 0.6, 0.4)$) {\small$\tau$};
        \node[oins_loc, fill=TealBlue!20] (s5) at ($(u1.south) + (-0.6,-0.4)$) {\small$\tau$};
        \node[oins_loc, fill=TealBlue!20] (s6) at ($(u1.south) + (-0.2,-0.4)$) {\small$\tau$};
        \node[oins_loc, fill=TealBlue!20] (s7) at ($(u1.south) + ( 0.2,-0.4)$) {\small$\tau$};
        \node[oins_loc, fill=TealBlue!20] (s8) at ($(u1.south) + ( 0.6,-0.4)$) {\small$\tau$};
        
        \draw[thick] (s1) -- ($(u1.north) + (-0.6,0)$);
        \draw[thick] (s2) -- ($(u1.north) + (-0.2,0)$);
        \draw[thick] (s3) -- ($(u1.north) + ( 0.2,0)$);
        \draw[thick] (s4) -- ($(u1.north) + ( 0.6,0)$);
        \draw[thick] (s5) -- ($(u1.south) + (-0.6,0)$);
        \draw[thick] (s6) -- ($(u1.south) + (-0.2,0)$);
        \draw[thick] (s7) -- ($(u1.south) + ( 0.2,0)$);
        \draw[thick] (s8) -- ($(u1.south) + ( 0.6,0)$);
        \draw[thick] ($(s1) + (0, 0.4)$) -- (s1);
        \draw[thick] ($(s2) + (0, 0.4)$) -- (s2);
        \draw[thick] ($(s3) + (0, 0.4)$) -- (s3);
        \draw[thick] ($(s4) + (0, 0.4)$) -- (s4);
        \draw[thick] ($(s5) + (0,-0.4)$) -- (s5);
        \draw[thick] ($(s6) + (0,-0.4)$) -- (s6);
        \draw[thick] ($(s7) + (0,-0.4)$) -- (s7);
        \draw[thick] ($(s8) + (0,-0.4)$) -- (s8);
    } = 
    \tikz[baseline=-0.5ex]{
        \tikzset{
            wten/.style={
                draw,
                fill=BurntOrange!20,
                inner sep = 0,
                minimum width = 1.6cm,
                minimum height = 0.6cm
        }}
        \node[wten] (u1) at (0,0) {$u_j$};
        \draw[thick] ($(u1.south) + (-0.6,-0.4)$) -- ($(u1.south) + (-0.6,0)$);
        \draw[thick] ($(u1.south) + (-0.2,-0.4)$) -- ($(u1.south) + (-0.2,0)$);
        \draw[thick] ($(u1.south) + ( 0.2,-0.4)$) -- ($(u1.south) + ( 0.2,0)$);
        \draw[thick] ($(u1.south) + ( 0.6,-0.4)$) -- ($(u1.south) + ( 0.6,0)$);
        \draw[thick] ($(u1.north) + (-0.6, 0.4)$) -- ($(u1.north) + (-0.6,0)$);
        \draw[thick] ($(u1.north) + (-0.2, 0.4)$) -- ($(u1.north) + (-0.2,0)$);
        \draw[thick] ($(u1.north) + ( 0.2, 0.4)$) -- ($(u1.north) + ( 0.2,0)$);
        \draw[thick] ($(u1.north) + ( 0.6, 0.4)$) -- ($(u1.north) + ( 0.6,0)$);
    }
\]

By contracting this circuit applied to the seed operator of Eq.~\eqref{eq:seed_operatator_general} without any truncation, the resulting MPO is an exact MPU with a bond dimension of~\cite{Wahl2017}
\begin{equation}\label{eq:bond_dim_ansatz}
    \chi = n^{2k} \eqstop
\end{equation}
This allows the expressivity of the ansatz to be controlled by tuning the block size $k$ of the local unitaries.

Using this ansatz with a fixed seed operator $\Gamma_0$, the commutator functional from Eq.~\eqref{eq:commutator_functional} becomes a function of the unitary circuit
\[
\costfn(\szm) = \costfn(\unitary, \unitary^\dag)\eqstop
\]
The minimization of this functional can be performed directly on the manifold of the local unitaries using Riemannian optimization schemes, as explained in~\cite{Hauru2021}.
In short, Riemannian optimization uses the bare gradients $\nabla_{u_{j}^{(\dag)}} \costfn$, evaluated by treating the unitary $u_{j}^{(\dag)}$ as a normal matrix, and projects them to the tangent space of the manifold of unitary matrices.
The next update step is then determined by varying the unitary along a path in this manifold in the direction of the projected gradient.
The bare gradients are calculated using automatic differentiation~\cite{Nocedal2006}, and the cost function is minimized with respect to all unitaries $u_j$ in the circuit simultaneously.
Note that this strategy becomes very expensive for large block sizes $k$ of the local unitaries, and it is preferable to sweep through the circuit and optimize each local tensor individually by fixing the rest~\cite{Pollmann2016, Wahl2017}.
At the beginning of the optimization, the unitaries $u_j$ are initialized close to the identity using
\begin{equation}\label{eq:initialization}
    u_j = \exp(i\alpha V_j)
\end{equation}
where $V_j$ is a random Hermitian operator of block size $k$ acting on the sites $\lbrace (j-1)k+1, (j-1)k+2, jk\rbrace$ with unit norm that commutes with the parafermionic parity operator $\paraparity$, and $\alpha \ll 1$ is a small rotation angle.

\fi

\section{\label{sec:ising_case}Case  \texorpdfstring{$\MakeLowercase{n}=2$}{n=2}}

\if\includecasetwo1

We first test our ansatz by setting $n=2$.
In this case, the parafermions defined in Eqs.~\eqref{eq:parafermions_def_onsite} and~\eqref{eq:parafermions_def_offsite} become Majorana operators.
A typical model that is known to exhibit an extended parameter region hosting a topological phase with fractionalized MZMs is the Kitaev-Hubbard chain~\cite{Hassler2012, Katsura2015, Mahyaeh2020}.
In the spin language, the two spin operators defined by the generalized Jordan-Wigner transformation of Eq.~\eqref{eq:jw_traffo_gen} can be represented by the two Pauli matrices
\begin{equation}\label{eq:pauli_matrices}
    \tau_j = \paulz{j} = \begin{pmatrix}
        1 & 0\\ 0 & -1
    \end{pmatrix}
    \eqcmm\ 
    \sigma_j = \paulx{j} = \begin{pmatrix}
        0 & 1\\ 1 & 0
    \end{pmatrix}\eqcmm
\end{equation}
and the Kitaev-Hubbard chain becomes
\begin{equation}\label{eq:kitaev_hubbard_chain}
    H_{\text{KH}} = \sum_{j=1}^{N-1} -w \paulx{j}\paulx{j+1} + U\paulz{j}\paulz{j+1} - \frac{\mu}{2}\sum_{j=1}^N \paulz{j}\eqstop
\end{equation}
%
%

\subsection{\label{subsec:ex_lim}Integrable limits}

The Kitaev-Hubbard model has two exactly solvable lines: $U = 0$ and $\mu = 0$.
For $U=0$, the model becomes the non-interacting Kitaev chain~\cite{Kitaev2001}, and it can be shown to possess a SZM inside the topological phase, i.e.,  $\mu < 2w$.
In the spin language, the SZM can be represented as an MPO of bond dimension $\chi = 2$ defined by the tensors
\begin{equation}\label{eq:kitaev_szm}
    W^j =
    \begin{pmatrix}
        \frac{\mu}{2w} \paulz{} & \mathcal{N} \paulx{} \\
        0 & \unit
    \end{pmatrix}\eqstop
\end{equation}
Here, the matrix describes the action of the tensor in the horizontal direction, the operators in the matrix define the vertical action, and $\mathcal{N}$ is an overall normalization constant that ensures that $\szm^2 = \unit$.

In the other limit, $\mu = 0$, it was found by Fendley~\cite{Fendley2016} that there exists an SZM for $U < w$.
This time, the SZM can be represented as an MPO of bond dimension $\chi = 3$ and tensors
\begin{equation}\label{eq:fendley_szm}
    W^j =
    \begin{pmatrix}
        0 & \pauly{} & \mathcal{N} \paulx{} \\
        -\frac{U}{w} \pauly{} & 0 & 0 \\
        0 & 0 & \unit
    \end{pmatrix}\eqstop
\end{equation}

Based on Eq.~\eqref{eq:bond_dim_ansatz}, the tensor network ansatz in Eq.~\eqref{eq:unitary_ansatz} should reproduce these exact MPOs with a block size of $k = 2$.
A direct calculation, see Appendix~\ref{app:derivation_exactly_solvable_limits} for the details, shows that this is true, by choosing the seed operator as the Pauli-X operator, i.e., 
\[
\szm_0 = \paulx{1}\eqstop
\]

Focusing first on the case $U = 0$, the MPO in Eq.~\eqref{eq:kitaev_szm} can be reproduced by choosing the parametrization of the two-body unitaries $u_j$ acting on site $(j,j+1)$ as
\begin{equation}\label{eq:parametrization_kitaev}
    u_j = e^{i\alpha_j \pauly{j}\paulx{j+1}} = \cos(\alpha_j)\unit + i\sin(\alpha_j) \pauly{j}\paulx{j+1}
\end{equation}
where the rotation angle is defined by
\begin{equation}\label{eq:rotation_angle_kitaev}
    \tan(2\alpha_j)^2 = \frac{\gamma^2}{1 - \gamma^2}\left( 1 - \gamma^{2(N - j)}\right)\eqcmm\ \gamma = \frac{\mu}{2w}\eqstop
\end{equation}

In the other case of $\mu = 0$, the circuit requires an alternating structure of unitaries acting differently on the odd-even index pairs $(2k-1, 2k)$ and the even-odd index pairs $(2k, 2k+1)$.
Let $u_{2k-1}$ be the unitaries in the ladder that act on the odd-even pairs, and $u_{2k}$ the unitaries acting on the even-odd pairs, the two different rotations are defined by
\begin{equation}\label{eq:parametrization_fendley}
\begin{split}
     u_{2k-1} &= e^{i\beta_k \paulz{2k-1} \paulz{2k}} = \cos(\beta_k)\unit + i\sin(\beta_k) \paulz{k} \paulz{k+1} \\
    u_{2k} &= e^{i\frac{\pi}{4} \paulx{2k} \paulx{2k+1}} = \frac{1}{\sqrt{2}}\left(\unit + i \paulx{k} \paulx{k+1} \right)\eqstop
\end{split}
\end{equation}
The angles $\beta_k$ with $1\le k < \left\lfloor \frac{N+1}{2}\right\rfloor$ defining the rotation on the odd-even index pairs are fixed by the set of equations
\begin{equation}\label{eq:rotation_angle_fendley}
    \tan(2\beta_k)^2 = \frac{\gamma^2}{1 - \gamma^2} \left(
        1 - \gamma^{2\left(\left\lfloor \frac{N+1}{2}\right\rfloor-j\right)}
    \right) \eqcmm \ \gamma = \frac{U}{w}\eqstop
\end{equation}
In both cases, the rotation angles in Eqs.~\eqref{eq:rotation_angle_kitaev} and~\eqref{eq:rotation_angle_kitaev} are only weakly dependent on the position for $j,k\ll N$.

\begin{figure}[thb] 
    \centering
    \includegraphics[width = 0.48\textwidth]{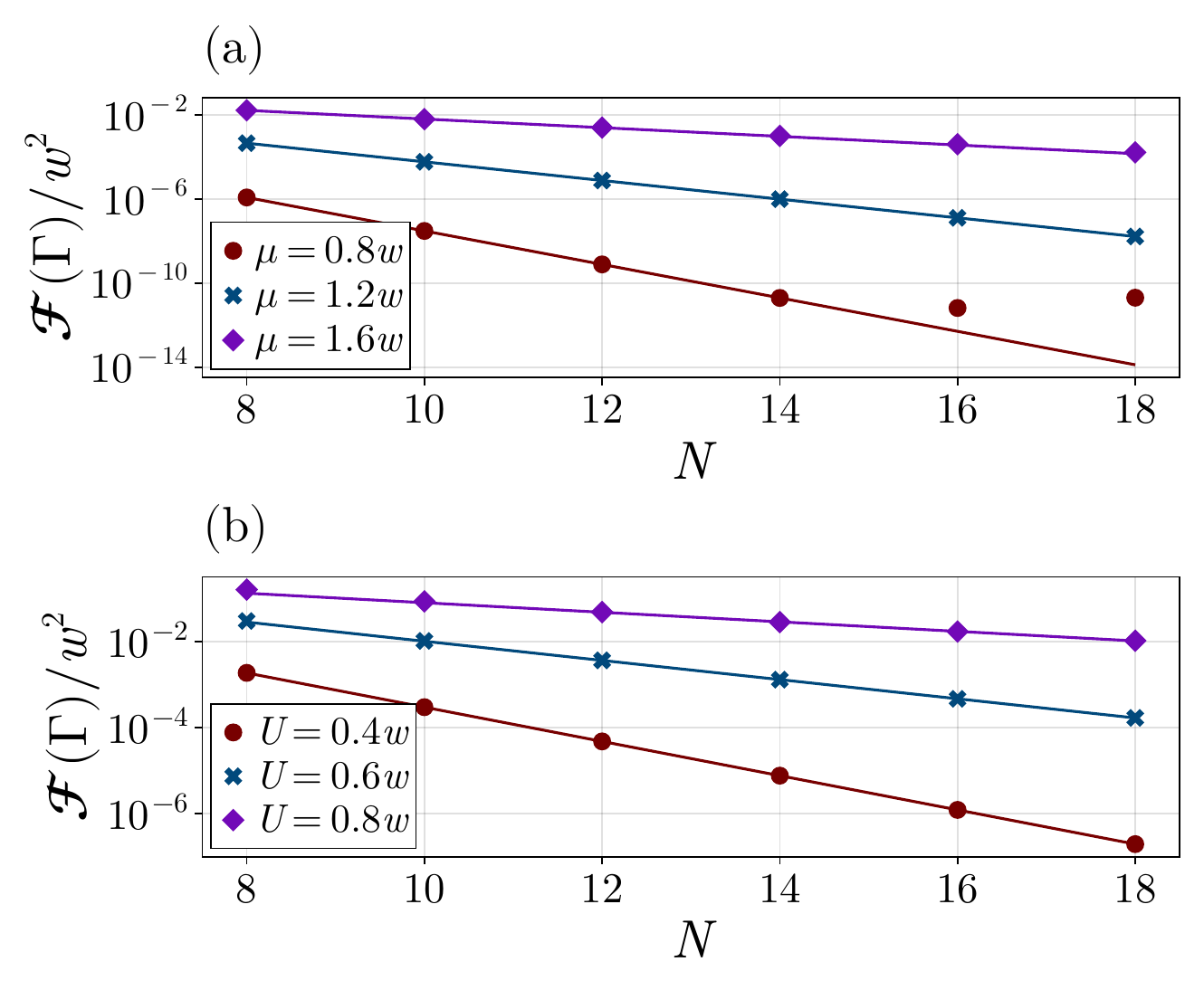}
    \caption{
        Commutator functional, Eq.~\eqref{eq:commutator_functional}, against the system size for $U = 0$ (panel (a)) and $\mu = 0$ (panel (b)).
        In panel (a) for $\mu = 0.8$, the commutator functional is limited by the sensitivity threshold set by the gradient norms.
        This results in the saturation of the obtained values.
        The lines are obtained by fitting the points with the function $f(N) = f_0 e^{-2\lambda_{\mathrm{fit}} N }$, with the results for $\lambda_{\mathrm{fit}}$ reported in Tables~\ref{tab:fits_localization_length_a} and~\ref{tab:fits_localization_length_b}.
    }\label{fig:exact_points_energy_functional}
\end{figure}
\begin{table}[h!]
\centering
    \begin{tabular}{| c | c | c | } 
         \hline
         $\mu/w$ & $\lambda_{\mathrm{exact}}$ & $\lambda_{\mathrm{fit}}$ \\
         \hline
         $0.8$ & $0.91629$ & $0.916283(5)$\\ 
         $1.2$ & $0.5108$ & $0.5111(1)$ \\ 
$1.6$ & $0.223$ & $0.237(3)$ \\ 
         \hline
        \end{tabular}
    \caption{Extracted localization length of the SZM for $U = 0$ by fitting the exponential decay of the commutator functional from Fig.~\ref{fig:exact_points_energy_functional}(a).
    }\label{tab:fits_localization_length_a}
\end{table}
\begin{table}[h!]
\centering
    \begin{tabular}{| c | c | c | } 
        \hline
        $U/w$ & $\lambda_{\mathrm{exact}}$ & $\lambda_{\mathrm{fit}}$ \\
        \hline
        $0.4$ & $0.458145$ & $0.458158(5)$\\ 
        $0.6$ & $0.2554$ & $0.2565(3)$ \\ 
        $0.8$ & $0.111$ & $0.127(3)$ \\ 
        \hline
    \end{tabular}
    \caption{Extracted localization length of the SZM for $\mu = 0$ by fitting the exponential decay of the commutator functional from Fig.~\ref{fig:exact_points_energy_functional}(b).
    }\label{tab:fits_localization_length_b}
\end{table}

After verifying that our ansatz can represent the correct result analytically, we checked whether the variational minimization converged to the same result.
The numerically obtained commutator functional $\costfn(\unitary,\unitary^\dag)$ of Eq.~\eqref{eq:commutator_functional} is plotted against the system size $N$ in Figure~\ref{fig:exact_points_energy_functional} for various parameters.
In this figure, panel (a) shows the case of $U=0$ and panel (b) presents the results for $\mu = 0$.
In both instances, the unitary circuit with a block size of $k=2$ was variationally minimized until the norms of the gradients fell below $10^{-10}$.
This establishes a natural cutoff for the precision of the commutator functional at the tenth digit.
At finite system size, it is expected that the commutator functional $\costfn(\unitary,\unitary^\dag)$ vanishes exponentially with the length of the chain as 
\begin{equation}\label{eq:commutator_functional_exp_decay}
    \costfn(\szm) \sim e^ {-2\lambda N} \, ,
\end{equation}
where $\lambda$ defines the localization length of the SZM.
From the exact form of the SZM, the localization length can be found to be
\begin{equation}\label{eq:correlation_length_exact}
    \lambda_{\text{exact}} = 
    \begin{cases}
        -\log\left(\frac{\mu}{2w}\right) \quad &\text{for } U = 0\\ 
        -\frac{1}{2}\log\left(\frac{U}{w}\right) \quad &\text{for } \mu = 0\\ 
    \end{cases}\eqstop
\end{equation}
For all parameters considered here, the commutator functional indeed vanishes exponentially with the system size, following Eq.~\eqref{eq:commutator_functional_exp_decay}. 
The localization lengths obtained by fitting the exponential decay are comparable to the exact values from Eq.~\eqref{eq:correlation_length_exact} and are presented in Table~\ref{tab:fits_localization_length_a} for $U = 0$ and Table~\ref{tab:fits_localization_length_b} for $\mu = 0$.
In the cases $\mu = 1.6w$ and $U = 0.8w$ the difference between the fit and the predicted values is the largest compared to the other cases.
We want to stress that this difference is not due to the precision of the method but comes from finite size effects.
The exact values for $\lambda$ are obtained from the exact solution of the edge mode, which only has an exponentially localized component.
Numerically, the edge mode will hybridize with the other end of the chain and the solution will slightly differ from the exact solution.
This is particularly expected for parameters where the edge mode is less localized.
We therefore expect the difference between the exact and predicted values to vanish for larger chains as the hybridization will become less important.

Next, we examine the operator distance
\begin{equation}\label{eq:infidelity}
    d(\hat{O}, \hat{K}) = \sqrt{\braket{
        \left(\hat{O} - \hat{K}\right)^\dag
        \left(\hat{O} - \hat{K}\right)
    }}
\end{equation}
between our variational optimal solution and the exact tensor networks from Eqs.~\eqref{eq:kitaev_szm} and~\eqref{eq:fendley_szm}.
This distance is shown in Fig.~\ref{fig:exact_points_overlap}, where panel (a) shows the case $U=0$ and panel (b) shows the case $\mu = 0$.
For all parameters, the distance exponentially vanishes with respect to the system size, up to the precision determined by the gradient convergence.
Performing the same analysis with an ansatz containing unitaries of larger block size reproduces the same results within numerical precision.
This clearly demonstrates that the variational ansatz with unitaries of block size $k=2$ is sufficient to obtain the correct MPOs of the SZM in the integrable limits, both numerically and analytically.
\begin{figure}[htb!]
    \centering
    \includegraphics[width = 0.48\textwidth]{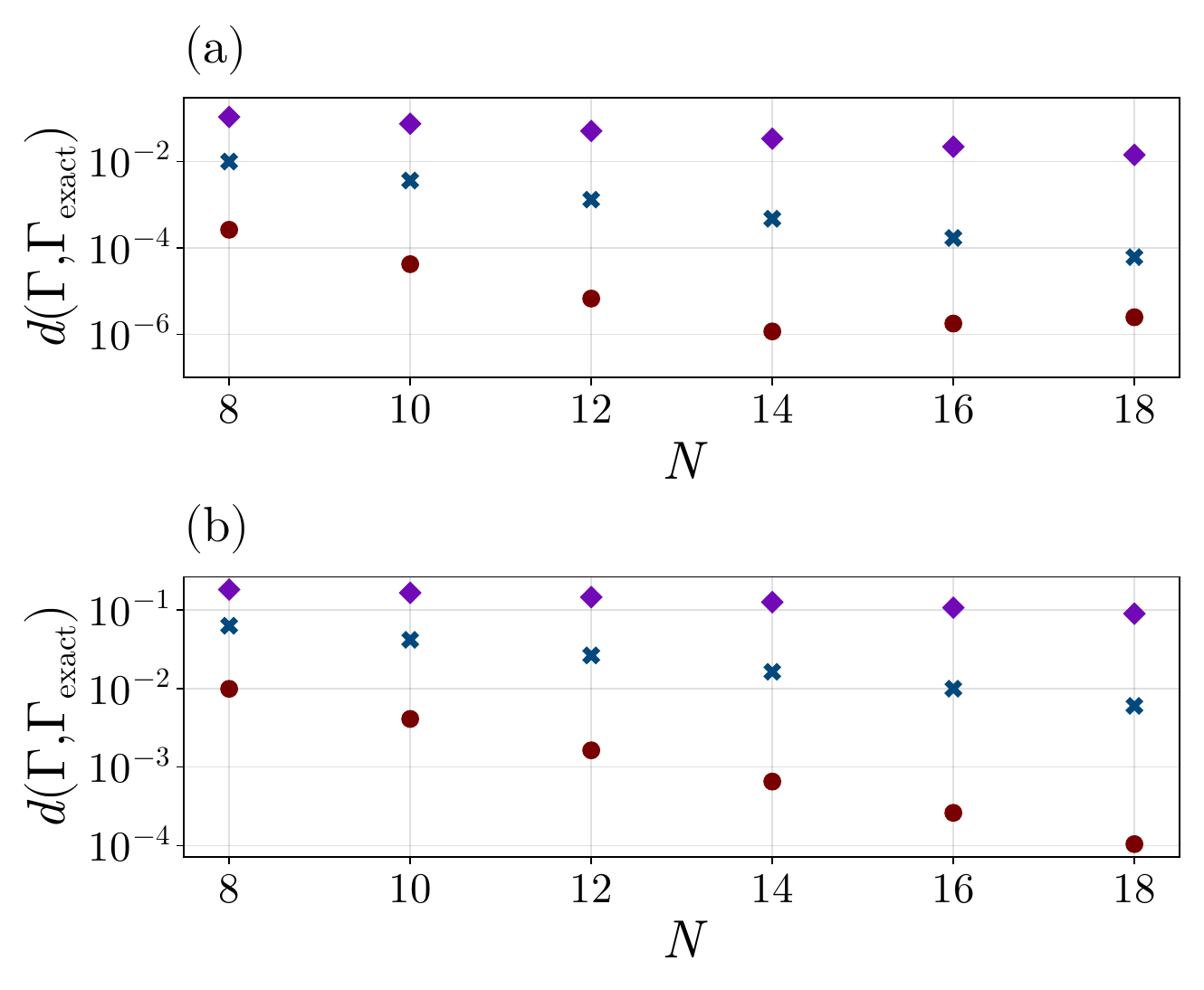}
    \caption{
    Distance from Eq.~\eqref{eq:infidelity} between the numerical optimal variational tensor network and the exact MPO with respect to the system size.
    Panel (a) shows the case of $U = 0$ with $\szm_{\mathrm{exact}}$ given by the MPO in Eq.~\eqref{eq:kitaev_szm}, and panel (b) the case of $\mu = 0$ where $\szm_{\mathrm{exact}}$ is given by the MPO from Eq.~\eqref{eq:fendley_szm}.
    The colors and markers represent the same parameters as in Fig.~\ref{fig:exact_points_energy_functional}.
    }\label{fig:exact_points_overlap}
\end{figure}

\subsection{\label{subsec:non-integrable}Generic points}

For generic points in the topological phase away from the integrable limits, the Kitaev-Hubbard chain in Eq.~\eqref{eq:kitaev_hubbard_chain} is expected to only possess an ASZM~\cite{Else2017,Parker2019,Kemp2020,Yates2020,Yates2020_B,Yeh2023,Olund2025, Tausendpfund2025}.
From the discussion in Section~\ref{sec:parafermion_aszm}, we expect to find an operator $\szm$ that minimizes the commutator functional $\costfn(\szm)$ and has a finite overlap with the edge operator $\paulx{1}$.

To test this expectation, we performed numerical minimization by varying the system size $N$ and the unitary block size $k$ appearing in the ansatz in Eq.~\eqref{eq:ladder_structure}.
For all the presented results, we minimized the commutator functional until the norms of the gradients fell below $10^{-6}$.
The threshold was chosen higher than in the integrable case, since the minimization procedure for generic points was more expensive.
Moreover, by lowering the threshold, we did not observe any qualitative difference.
%
%
Additionally, we fixed the chemical potential to $\mu = 0.2w$ and varied the interaction strength from $U=0.2w$ to $U=0.8w$.
With this set of parameters, we ensured that the Kitaev-Hubbard chain was always in the topological phase obeying a topological degenerate ground-state manifold~\cite{Mahyaeh2020}.

\begin{figure}[tbh]
    \centering
    \includegraphics[width=\linewidth]{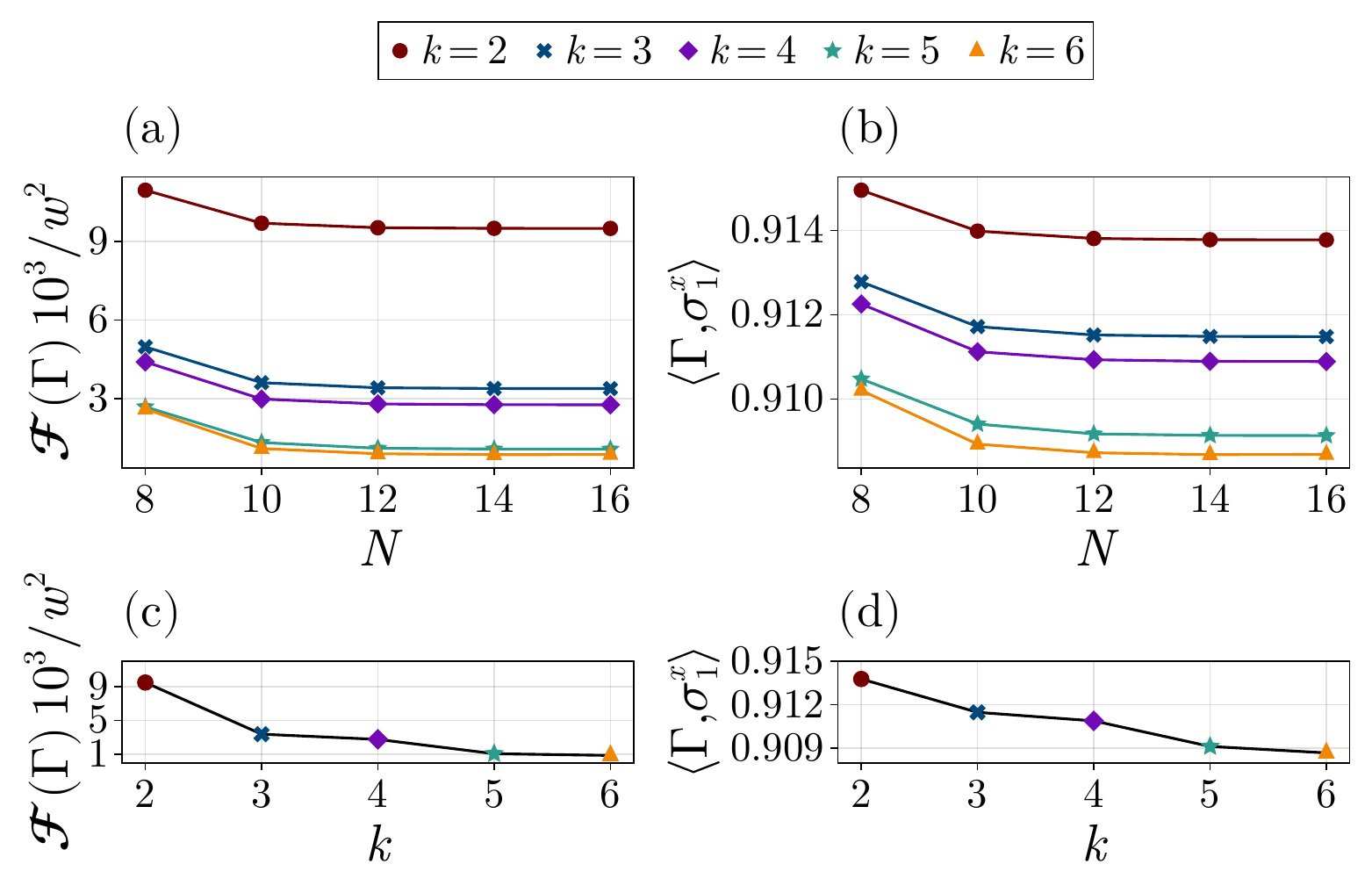}
    \caption{
        (a) Commutator functional obtained from numerical minimization at $\mu = 0.2w$ and $U=0.4w$ for an increasing block size of the unitaries $u_j$.
        (b) Overlap between the numerically obtained ASZM and the edge operator $\paulx{1}$ for the same parameter point as in panel (a).
        (c) Behavior of the commutator functional at a fixed system size of $N=16$ by increasing the block size $k$.
        (d) Same as (c), but showing the overlap.
    }
    \label{fig:energy_and_overlap_generic_point}
\end{figure}

Figure~\ref{fig:energy_and_overlap_generic_point}(a) presents the commutator functional of the numerically obtained $\szm(\unitary)$ for increasing system size, exemplified for $\mu = 0.2w$ and $U = 0.4w$.
For every fixed block size $k$, the commutator functional obtained from the minimization saturates with the system size at approximately $N = 12$.
Compared to the integrable limits shown in Fig.~\ref{fig:exact_points_energy_functional}, this saturation indicates that the operator found by our algorithm is indeed an ASZM with a finite lifetime.
By fixing the number of system sites, the commutator functional is reduced by increasing the block size $k$.
Figure~\ref{fig:energy_and_overlap_generic_point}(c) indicates that $\costfn(\szm)$ converges with respect to the block size $k$ to a value close to $\costfn(\szm) / w^2 \approx 10^{-3}$.
Because the unitary circuit preserves the locality of the seed operator, reaching a saturation value for a finite block size $k$ indicates that the ASZM is also localized to the edge of the system, exhibiting a finite penetration depth.

The overlap between the numerically determined $\szm(\unitary)$ and the edge operator $\paulx{1}$, as shown in Fig.~\ref{fig:energy_and_overlap_generic_point}(b) for $\mu = 0.2w$, $U = 0.4w$, further demonstrates the localization of the ASZM.
Similar to the commutator functional, the overlap saturates for a fixed block size with respect to the system size and decreases slightly with an increase in the block size, see also Fig.~\ref{fig:energy_and_overlap_generic_point}(d).
As a consequence, the edge operator can be expanded as
\[
    \paulx{1} = A \szm + \sqrt{1 - A^2}\,\tilde{\szm}\eqcmm \quad A = \braket{\szm,\paulx{1}}\eqcmm
\]
where $\tilde{\szm}$ is a remaining Majorana operator that anti-commutes with $\szm$.
Given that $\mathcal{F}(\paulx{1}) \sim w^2$ is valid for a generic point and $\szm$ possesses a small commutator functional, it follows that the residual operator must also satisfy $\mathcal{F}(\tilde{\szm})\sim w^2 \gg \mathcal{F}(\szm)$.
Consequently, the long-time properties of the edge excitation are defined by the ASZM $\szm$ and we expect that the lifetime of $\paulx{1}$, measured by the infinite temperature autocorrelation function~\cite{viswanath1994recursion, Parker2019_B, Nandy2025, Tausendpfund2025_B}
\[
 C(\paulx{1},t) = \braket{\paulx{1}e^{-itH_{\text{KH}}} \paulx{1}e^{itH_{\text{KH}}}} \approx A^2 e^{-t/\tau} \eqcmm
\]
is also of the order $w\tau \approx 10^{3}$.

\begin{figure}[htb!]
    \centering
    \includegraphics[width=0.85\linewidth]{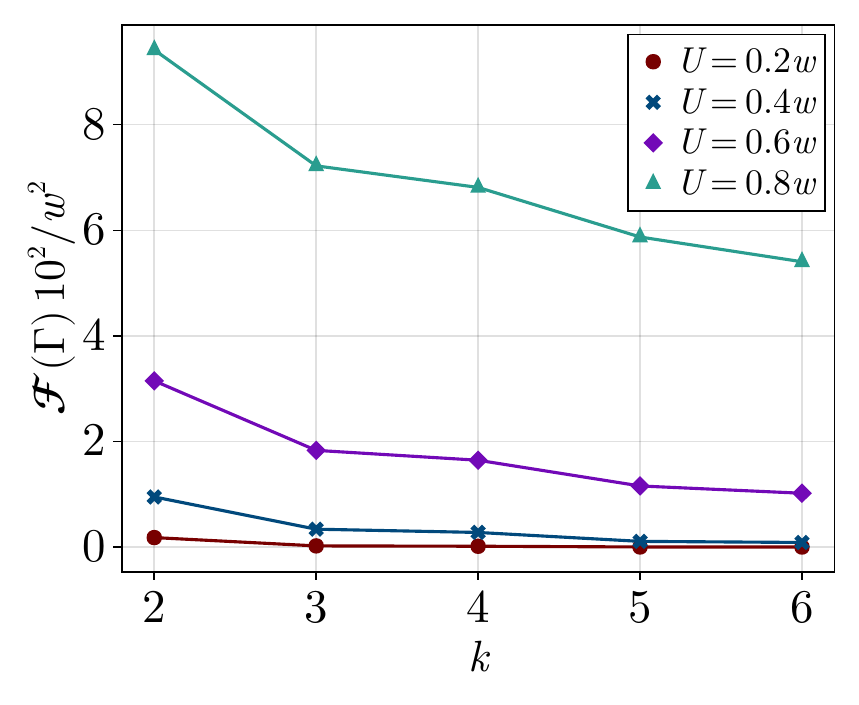}
    \caption{Commutator functional as a function of the block size $k$ and for different interaction strengths at $\mu = 0.2w$ and $N=16$.}
    \label{fig:convergence_interactions}
\end{figure}

The same convergence behavior in the commutator and overlap was also observed for other interaction strengths. 
This is explicitly shown in Fig.~\ref{fig:convergence_interactions}, where the commutator functional is plotted against the number of layers for various interaction strengths for a system with $N=16$ sites.
The final value of the saturation exhibits non-trivial behavior with respect to the interaction strength.
A similar non-trivial dependence on the interaction strength was also reported previously in~\cite{Tausendpfund2025}.
\begin{figure}[htb!]
    \centering
    \includegraphics[width=0.9\linewidth]{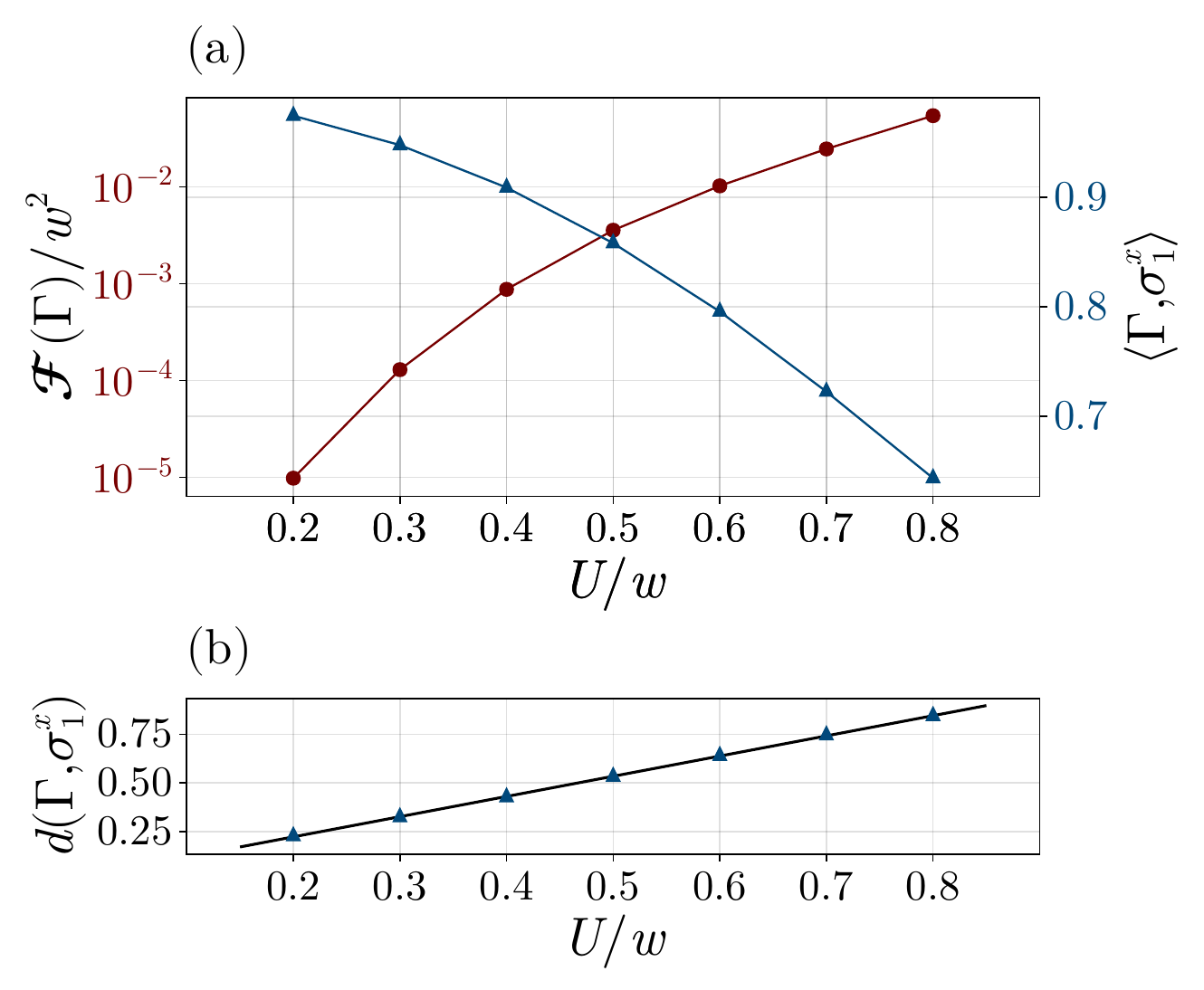}
    \caption{
    (a) Dependence of the commutator functional (red circles) and overlap with the seed operator $\szm_0$ (blue triangles) on the interaction strength for $\mu = 0.2w$ and a chain of $N = 16$ sites.
    (b) Distance, Eq.~\eqref{eq:infidelity}, of the numerically obtained ASZM from the seed operator $\paulx{1}$. 
    The black line is a linear fit of the form $f(U) = \delta U/w + d_0$ with $\delta = 1.037(6)$ and $d_0 = 0.0157(3)$.
    }\label{fig:energy_vs_interaction}
\end{figure}
In Figure~\ref{fig:energy_vs_interaction}(a), the dependence of the energy functional and the overlap with the edge operator $\paulx{1}$ on the interaction strength is shown for $N=16$ and block size $k = 6$.
%
Both quantities have a remarkable symmetry with respect to the interaction strength: while the logarithm commutator increases with increasing interaction strength, the overlap decreases.
Consequently, moving further away from the integrable limits gives rise to a more delocalized operator with a shorter lifetime.

From the Majorana properties of $\szm$ and $\paulx{1}$, we find that the distance in Eq.~\eqref{eq:infidelity} is connected to the overlap through
\[
 d(\szm,\paulx{1}) = \sqrt{2(1 - \braket{\szm,\paulx{1}})} \eqstop
\]
We observe that the distance depends linearly on the interaction strength
\[
     d(\szm,\paulx{1}) = \delta \frac{U}{w} + d_0\eqcmm
\]
as shown in Fig.~\ref{fig:energy_vs_interaction}(b) together with a linear fit.
Based on the mirror symmetry of Fig.~\ref{fig:energy_vs_interaction}(a), a similar form can be found for the commutator functional
\[
    \mathcal{F}(\szm) \sim w^2e^{-\frac{(\tilde{A}(1 - U/w) + \tilde{f}_0)^2}{2}}\eqcmm
\]
for the intermediate interactions considered in this work.
Consequentially, the lifetime of the ASZM approaches order $ w \tau \simeq 1$ exponentially fast, as $U$ approaches $w$
\[
    w\tau \sim e^{\tilde{A}^2(1 - U/w)^2 }\eqstop
\]

We emphasize that this functional behavior is incorrect for small values of $U$, as the commutator must vanish exponentially with the system size when $U=0$, as shown in Eq.~\eqref{eq:commutator_functional_exp_decay}.

\fi

\section{\label{sec:chiral_potts_model}Case \texorpdfstring{$\MakeLowercase{n}=3$}{n=3}}

\if\includecasethree1

\subsection{Model}

Next, we consider the case of $\Z_3$-parafermions.
In particular, we are interested in the model~\cite{Fendley2012}
\begin{equation}\label{eq:chiral_z3_parafermions_hamiltonian}
    \begin{split}
    H_3 &= -w \sum_{j=1}^{N-1} \left(\psi_j^{\dag}\chi_{j+1}^{\vdag}e^{-i\chiralangle} \omega + \hc\right)\\
    {}&-\frac{\mu}{2} \sum_{j=1}^{N} 
    \left(\chi_j^\dag \psi_{j}^{\vdag} e^{-i\chiralangle}\omega +\hc \right) \eqcmm
    \end{split}
\end{equation}
where $\psi_j$ and $\chi_j$ are the $\Z_3$ parafermions introduced in Sec.~\ref{sec:parafermion_aszm}.
Although the model in Eq.~\eqref{eq:chiral_z3_parafermions_hamiltonian} is quadratic in the parafermions, it is not exactly solvable because of the non-trivial commutation relations between the parafermions at distance, see Eq.~\eqref{eq:parafermions_def_offsite}, except at special points~\cite{Auyang1987,Vongehlen1985}.
Passing to the spin variables using the transformation~\eqref{eq:jw_traffo_gen}, this model becomes the chiral Potts model~\cite{Fendley2012}
\begin{equation}\label{eq:chiral_potts_model}
    \begin{split}
        H_3 &= -w \sum_{j=1}^{N-1}\left(
    	 \sigma^\dagger_{j} \sigma^{\vdag}_{j+1} e^{-i\chiralangle} + \sigma^{\vdag}_{j} \sigma^\dag_{j+1} e^{i\chiralangle}
    	 \right)\\
        {}&-\frac{\mu}{2} \sum_{j=1}^N \left(
    		\tau_j^\dag e^{-i\chiralangle} + \tau^{\vdag}_j e^{i\chiralangle}
    	\right)
    \end{split}
	 \eqstop
\end{equation}
This model, although particularly simple, has a rich phase diagram~\cite{Huse1981, Howes1983, Huse1983}, including a three-fold degenerate phase associated with boundary-localized parafermions~\cite{Fendley2012}.
The phase diagram of the model is expected to be symmetric in $\chiralangle = \pi/6$.~\cite{Fendley2012}
In particular, Fendley argued that this model should possess an SZM, at least at the special point of $\chiralangle = \pi/6$~\cite{Fendley2012}.
At this point, the Hamiltonian becomes symmetric around zero energy, $H = -H$, integrable, and shares many properties with the $U=0$ limit of the Kitaev-Hubbard chain from Eq.~\eqref{eq:kitaev_hubbard_chain}.
Furthermore, the leading order of the SZM is argued to be given by
\begin{equation}\label{eq:z3szm_leading_order}
    \szm = \mathcal{N}\left(\sigma_1 - i\frac{\mu}{4w\sin(3\chiralangle)} \hat{X} + \dots\right) \, ,
\end{equation}
where $\hat{X}$ is a linear combination of parafermion strings up to the second site, and $\mathcal{N}$ is an overall normalization constant that ensures $\szm^\dag\szm = \unit$. The rest of the operator is expected to be composed of larger parafermionic strings with higher orders of $\mu/(4w\sin(3\chiralangle))$~\cite{Fendley2012}.
This result indicates that an SZM exists for parameter combinations where
\begin{equation}\label{eq:z3_inequality}
    \left|\frac{\mu}{4w\sin(3\chiralangle)} \right| < 1\eqstop
\end{equation}

Similar to the Majorana case discussed in Sec.~\ref{sec:ising_case}, we perform a numerical minimization of our ansatz for different block sizes $k$.
Based on the leading-order contribution~\eqref{eq:z3szm_leading_order}, we set the seed operator to $\szm_0 = \sigma_1$ and variationally minimize the circuit until the gradients were below $10^{-6}$.

\subsection{Results}
We begin by discussion the general convergence behavior of the numerical minimization with respect to the system and block size.
Figure~\ref{fig:parafermions_energy_and_overlap_generic_point}(a) presents the commutator functional of the numerically obtained $\szm(\unitary)$ for the $n=3$ parafermions for increasing system size for the parameter point $\mu = w$ and $\chiralangle = \pi/12$.
This point is not close to the special point of $\chiralangle = \pi/6$ but, based on the inequality of Eq.~\eqref{eq:z3_inequality}, is expected to host a SZM:
\[
 \left|\frac{1}{4\sin(\pi/4)} \right| = 0.\overline{35} \eqstop
\]
Similar to the $n=2$ case, Fig.~\ref{fig:energy_and_overlap_generic_point}(a), we observe that the commutator functional converges for every fixed block size $k$ with respect to the system size, although this convergence already happens at a rather small system size of $N=6$.

\begin{figure}[t!]
    \centering
    \includegraphics[width=\linewidth]{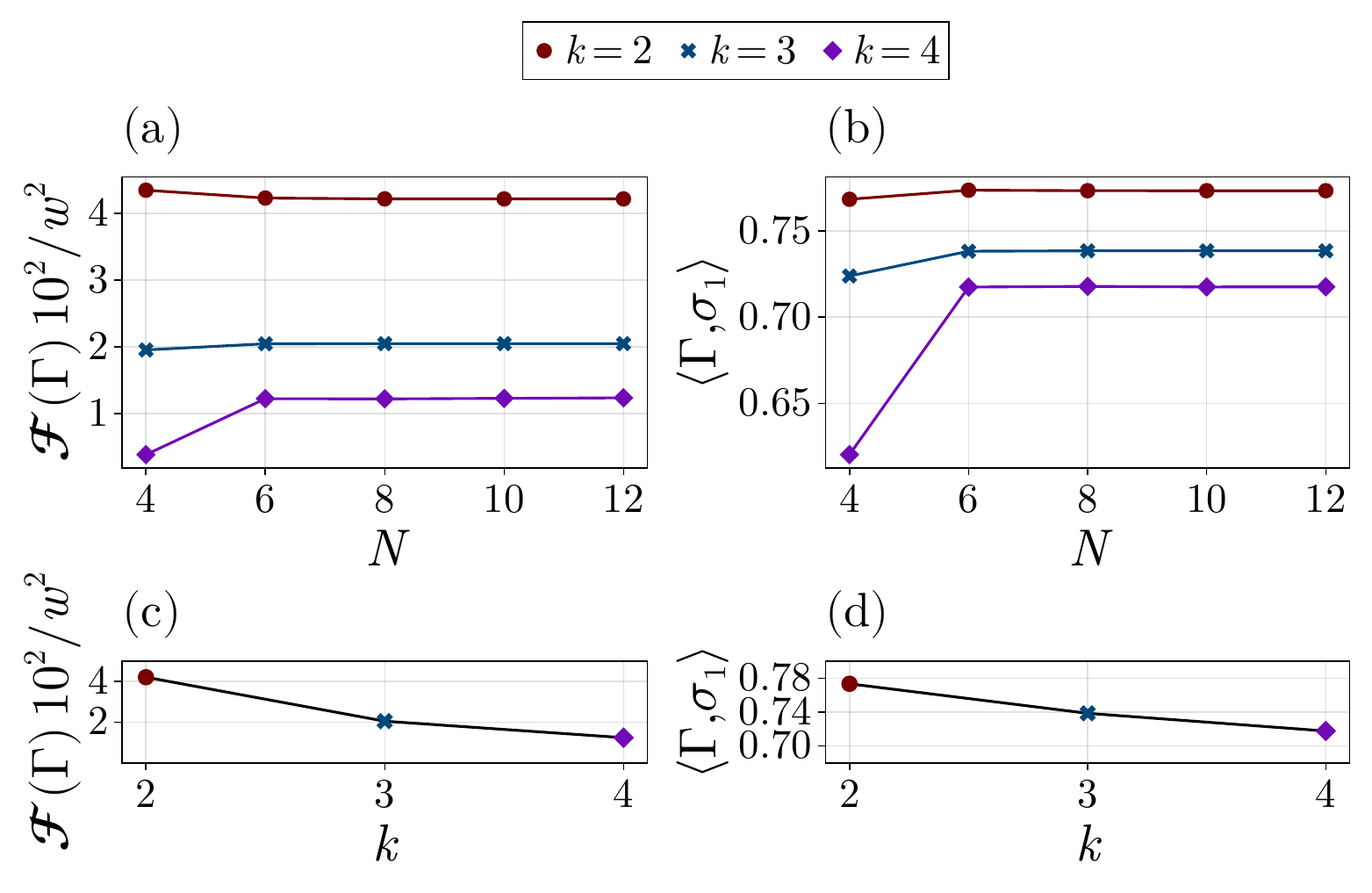}
    \caption{
        (a) Commutator functional obtained from numerical minimization at $\mu = w$ and $\chiralangle=\pi/12$ for an increasing block size of the unitaries $u_j$.
        (b) Overlap between the numerically obtained ASZM and the edge operator $\sigma_1$ for the same parameter point as in panel (a).
        (c) Behavior of the commutator functional at a fixed system size of $N=12$ by increasing the block size $k$.
        (d) Same as (c), but showing the overlap.
    }
    \label{fig:parafermions_energy_and_overlap_generic_point}
\end{figure}

Furthermore, by varying the block size $k$ while fixing the system size, we observe a converging behavior of the commutator functional.
This is shown in Fig.~\ref{fig:parafermions_energy_and_overlap_generic_point}(c) indicating a limiting value close to $\mathcal{F}(\szm)/w^2 \approx 10^{-2}$.
This behavior signals the appearance of an ASZM with a finite lifetime rather than an SZM, as predicted by the inequality from Eq.~\eqref{eq:z3_inequality}.

The overlap between the edge operator $\sigma_1$ and the numerically found ASZM $\szm(\mathcal{U})$ demonstrates the localization properties of ASZM.
Similar to the commutator functional, this overlap converges in system size, see Fig.~\ref{fig:parafermions_energy_and_overlap_generic_point}(b), and in the block size, see Fig.~\ref{fig:parafermions_energy_and_overlap_generic_point} (d).
\begin{figure}[htb!]
    \centering
    \includegraphics[width=0.85\linewidth]{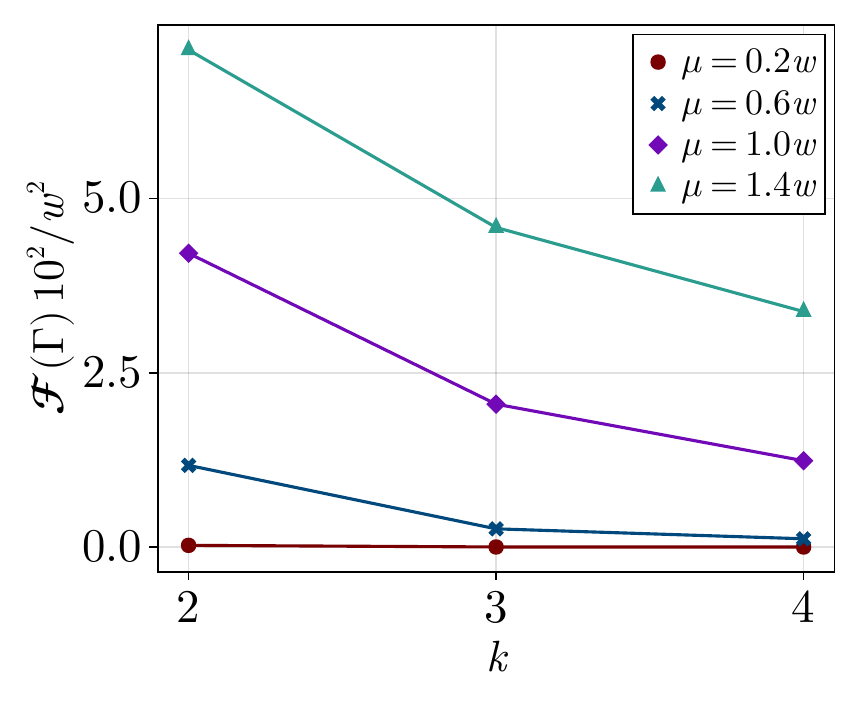}
    \caption{Commutator functional as a function of the block size $k$ and for different coupling strengths at $\chiralangle = \pi/12$ and $N=12$.}
\label{fig:parafermions_convergence_interactions}
\end{figure}
The same convergence properties are observed for a wide range of parameters as shown in Fig.\ref{fig:parafermions_convergence_interactions} for a fixed chiral angle of $\varphi = \pi/12$.

\begin{figure}[htb!]
    \centering
    \includegraphics[width=0.9\linewidth]{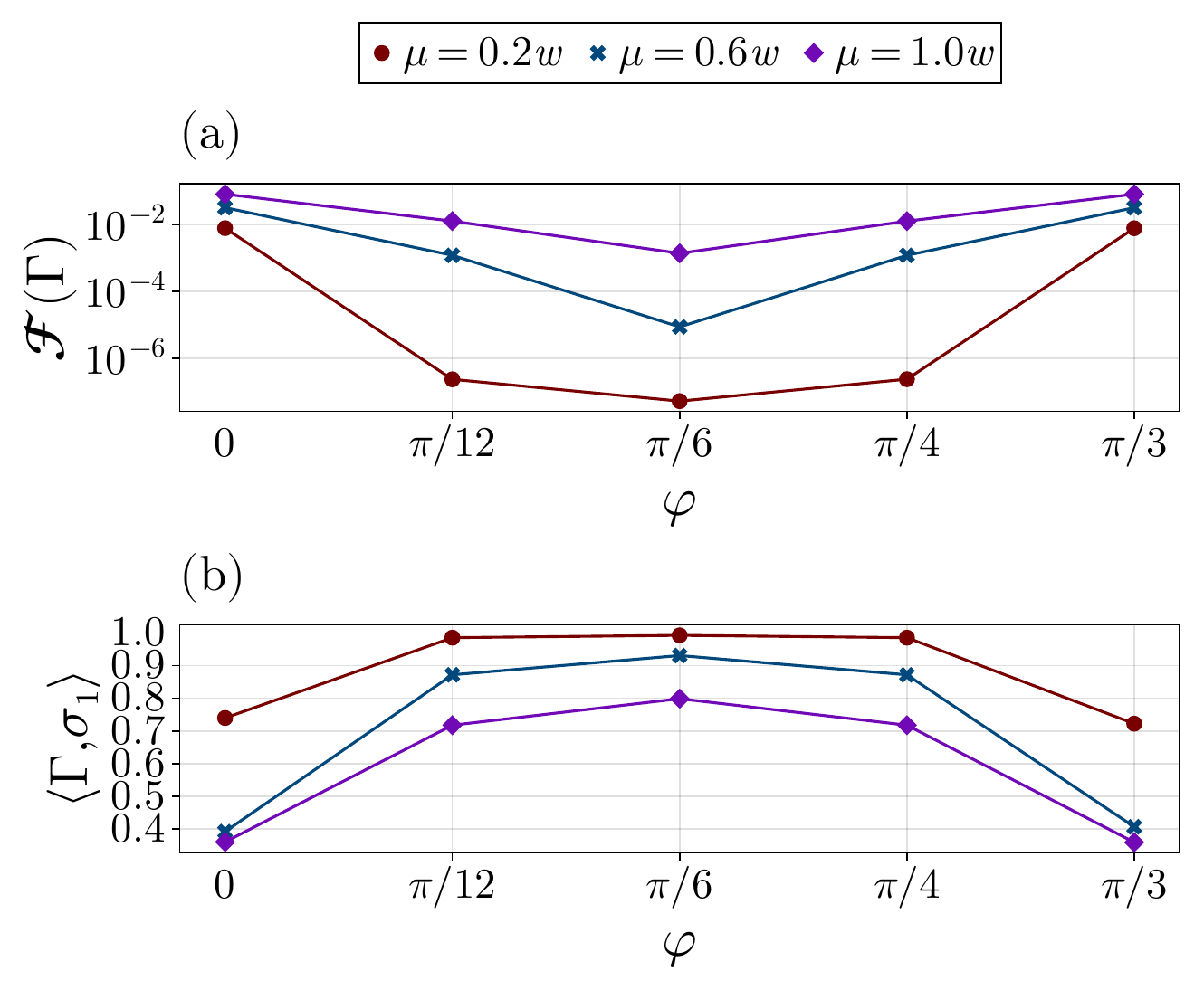}
    \caption{Commutator functional, panel (a), and overlap to the edge operator $\sigma_1$ for varying chiral angle while fixing the block size to $k=4$ and system size to $N=12$.}
\label{fig:parafermions_phi_sweep}
\end{figure}
Next we discuss the behavior of the commutator functional and overlap as a function of the chiral angle $\chiralangle$.
For this, we fix the system size to $N=12$ and the block size to $k=4$.
The results are shown for $\mu/w\in \lbrace0.2,\,0.4,0.6\rbrace$ in Fig.~\ref{fig:parafermions_phi_sweep}(a) for the commutator functional and Fig.~\ref{fig:parafermions_phi_sweep}(b) for the overlap.
The results show the expected symmetry around the special point of $\chiralangle = \pi/6$. 
All simulated points show a small commutator functional while simultaneously being localized to the edge as indicated by a finite overlap with the edge operator.
This indicates the emergence of a ASZM with prolonged lifetimes through the whole parameter set.

The commutator functional becomes smallest at the special point of $\chiralangle = \pi/6$, as expected.
For $\mu=0.2w$ we further observe that the commutator functional is at the order of the numerical threshold of the gradients at $10^{-6}$.
This is also true in the vicinity of $\chiralangle = \pi/6$.
As a conclusion, we expect the model to host a stable SZM with an infinite lifetime at this parameter point, as conjectured in previous works~\cite{Fendley2012}.

The value of the commutator functional is largest towards the edge values $\chiralangle=0,\pi/3$ for all values of $\mu$.
This is expected, as in these values of $\chiralangle$ maximize the left-site of the inequality in Eq.~\ref{eq:z3_inequality}.
However, as the commutator functional is still two orders of magnitude smaller than for a typical operator, we conclude that the model generically hosts ASZM in the full parameter space considered in this study.

\fi
\if\includecasesamplepic

\section{Figures Parafermions -- ToDo}
\begin{figure}[htb!]
    \centering
    \includegraphics[width=\linewidth]{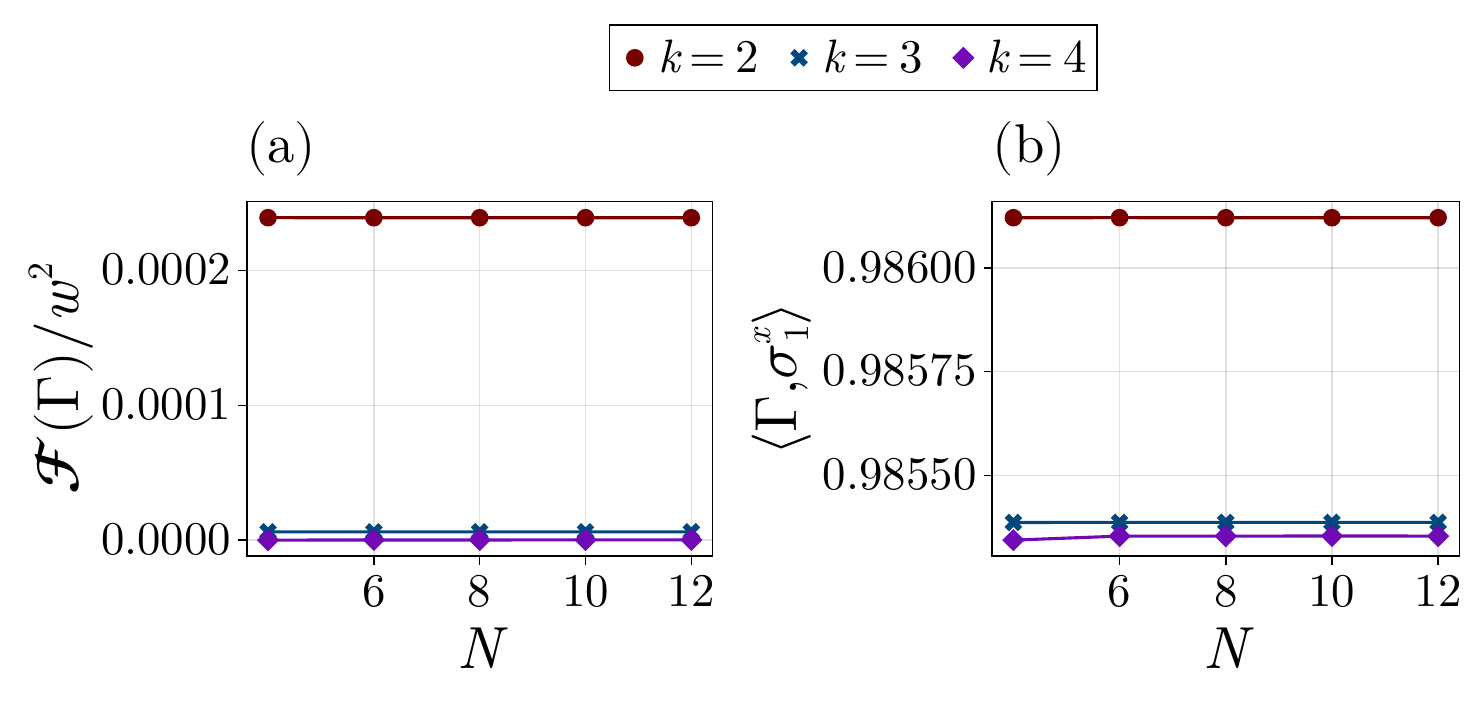}
    \caption{    
    Todo
    $\mu/2 = 0.1, 3\phi/pi = 2$
    }
\end{figure}%
\begin{figure}[htb!]
    \centering
    \includegraphics[width=\linewidth]{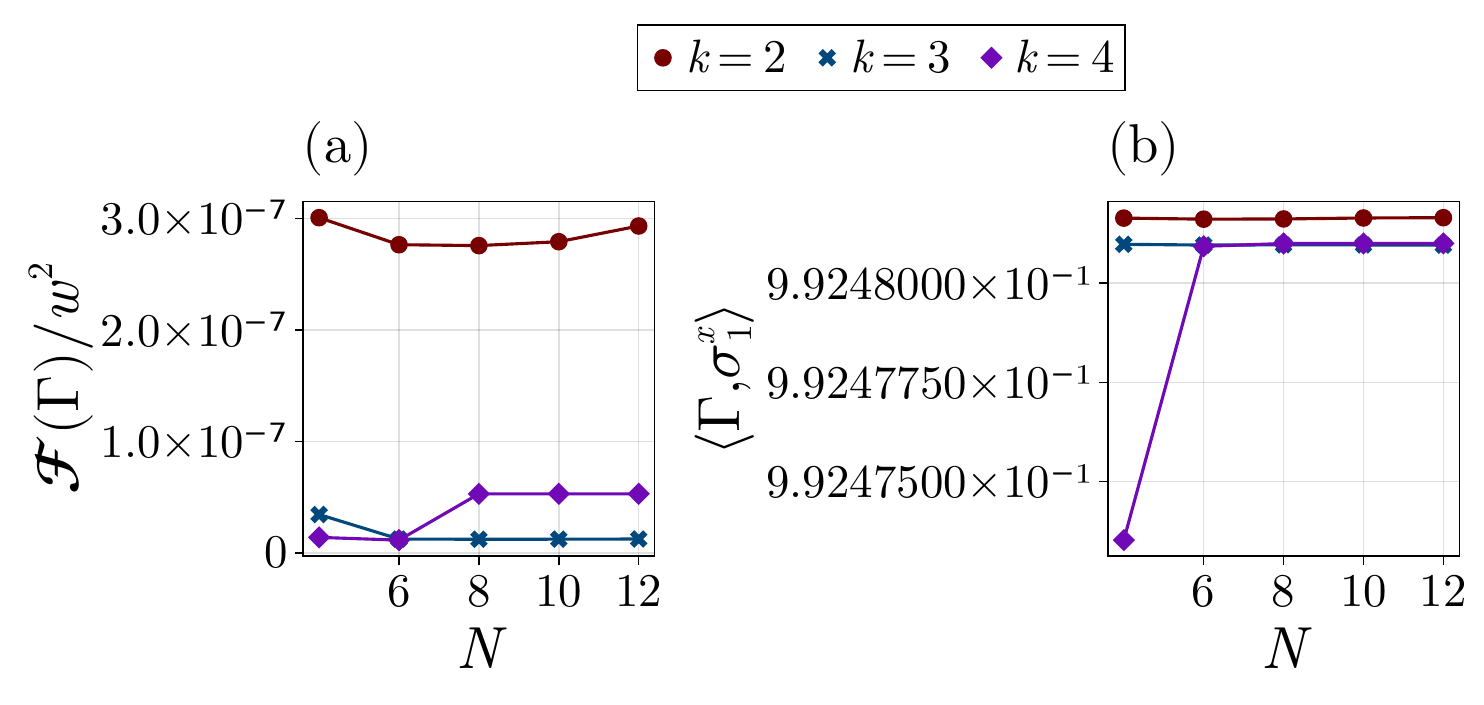}
    \caption{
    Todo
    $\mu/2 = 0.1,  3\phi/pi = 3$
    }
\end{figure}%
\begin{figure}[htb!]
    \centering
    \includegraphics[width=\linewidth]{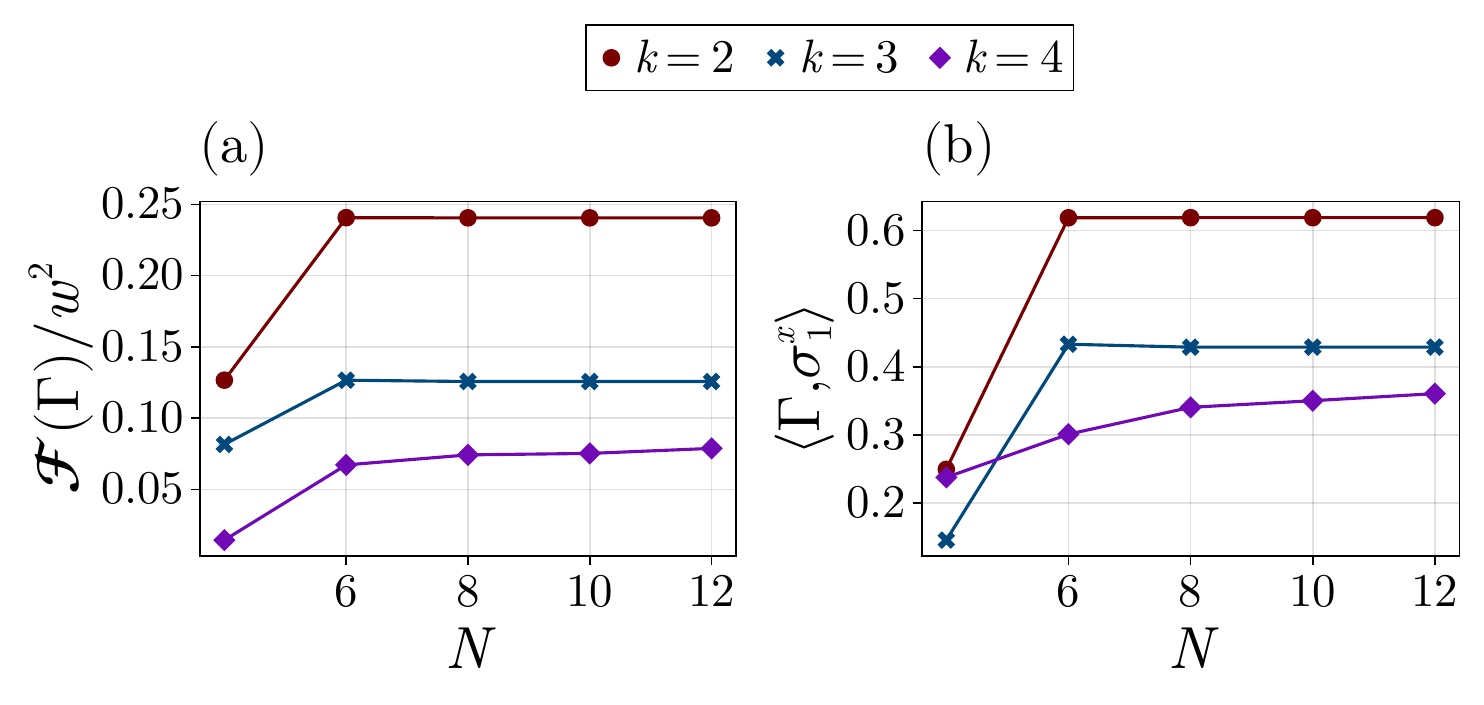}
    \caption{
    Todo
    $\mu/2 = 0.5, 3\phi/pi = 1$
    }
\end{figure}%
\begin{figure}[htb!]
    \centering
    \includegraphics[width=\linewidth]{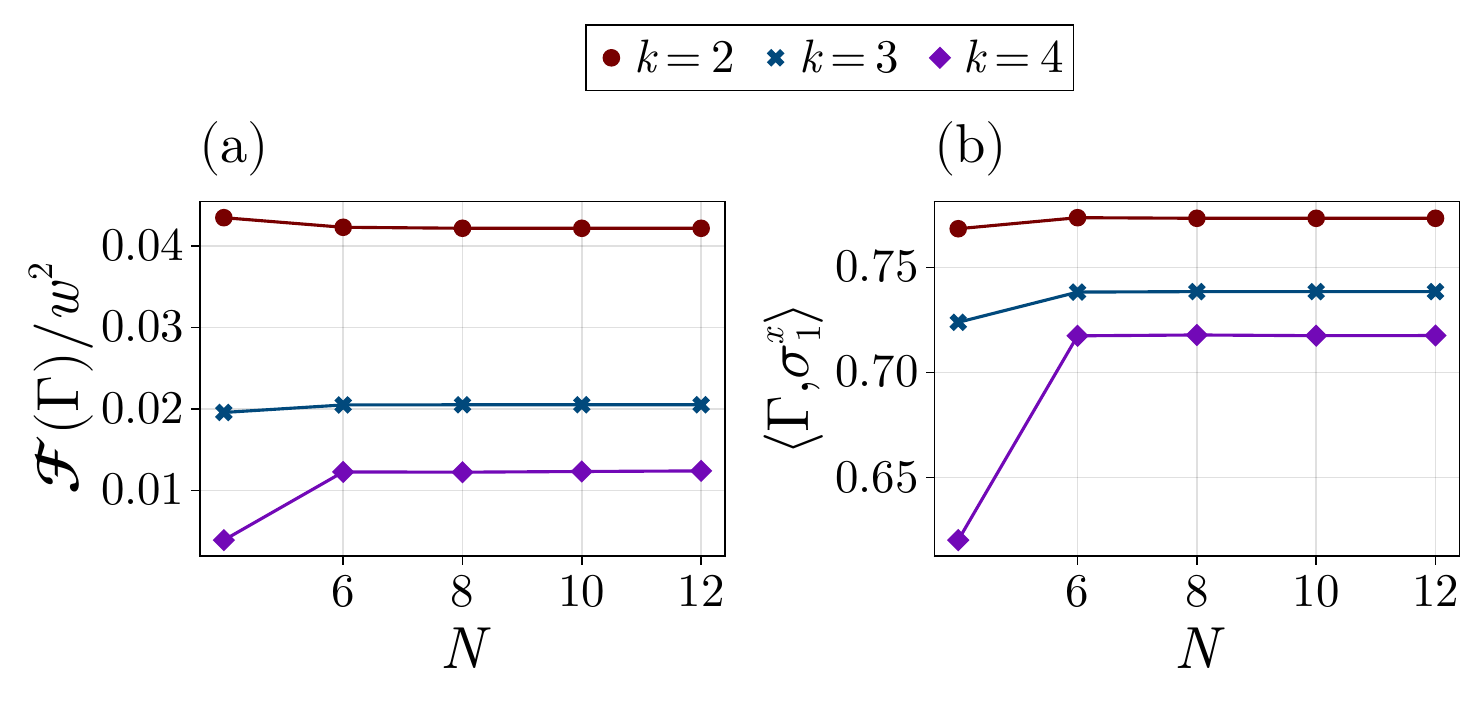}
    \caption{
    Todo
    $\mu/2 = 0.5, 3\phi/pi = 2$
    }
\end{figure}%
\begin{figure}[htb!]
    \centering
    \includegraphics[width=\linewidth]{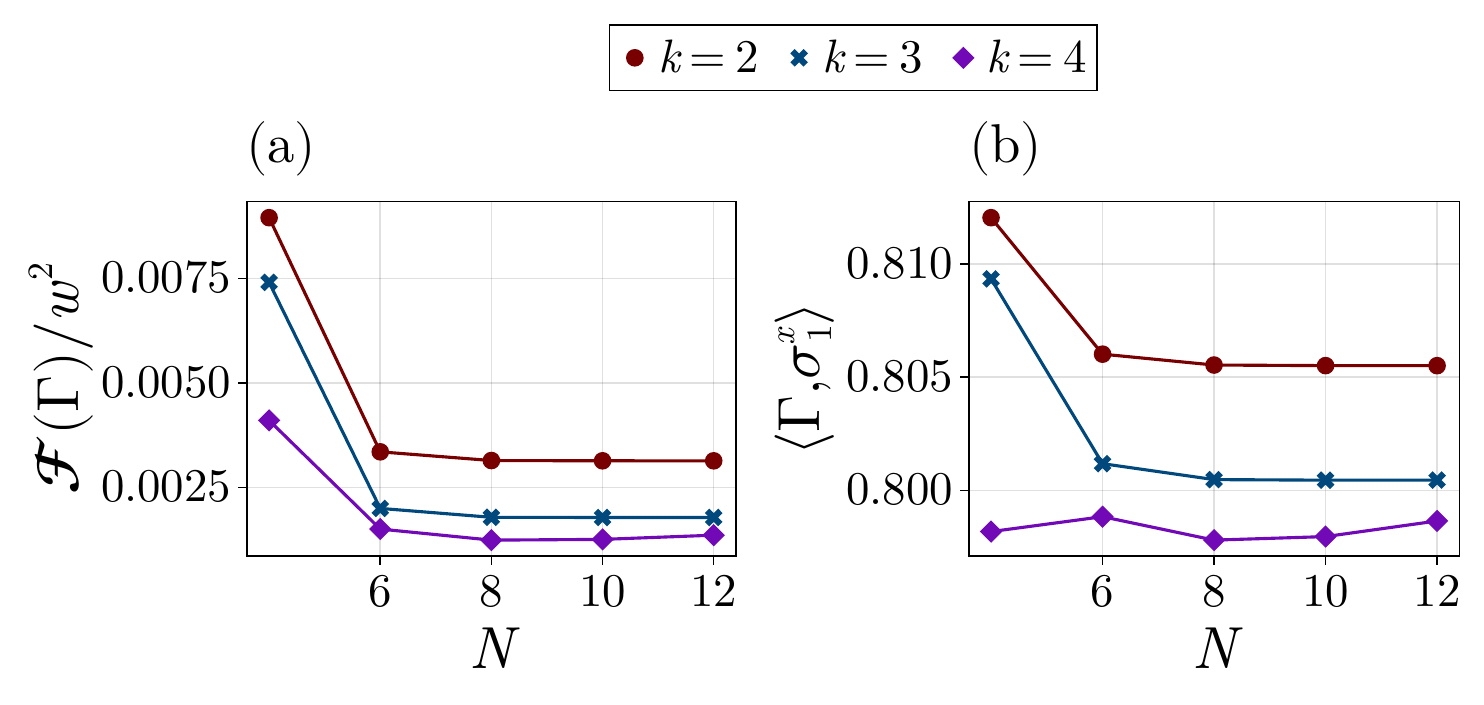}
    \caption{
    Todo
    $\mu/2 = 0.5, 3\phi/pi = 3$
    }
\end{figure}%
\begin{figure}[htb!]
    \centering
    \includegraphics[width=0.5\linewidth]{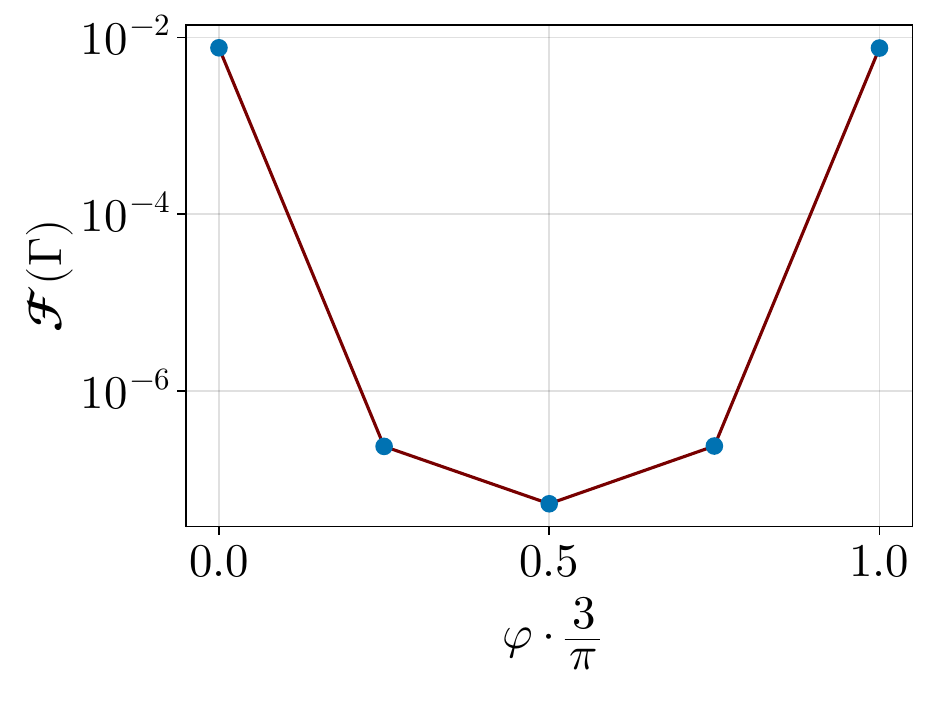}
    \caption{
    Todo
    $\mu/2 = 0.1, L=12$
    }
\end{figure}%
\begin{figure}[htb!]
    \centering
    \includegraphics[width=0.5\linewidth]{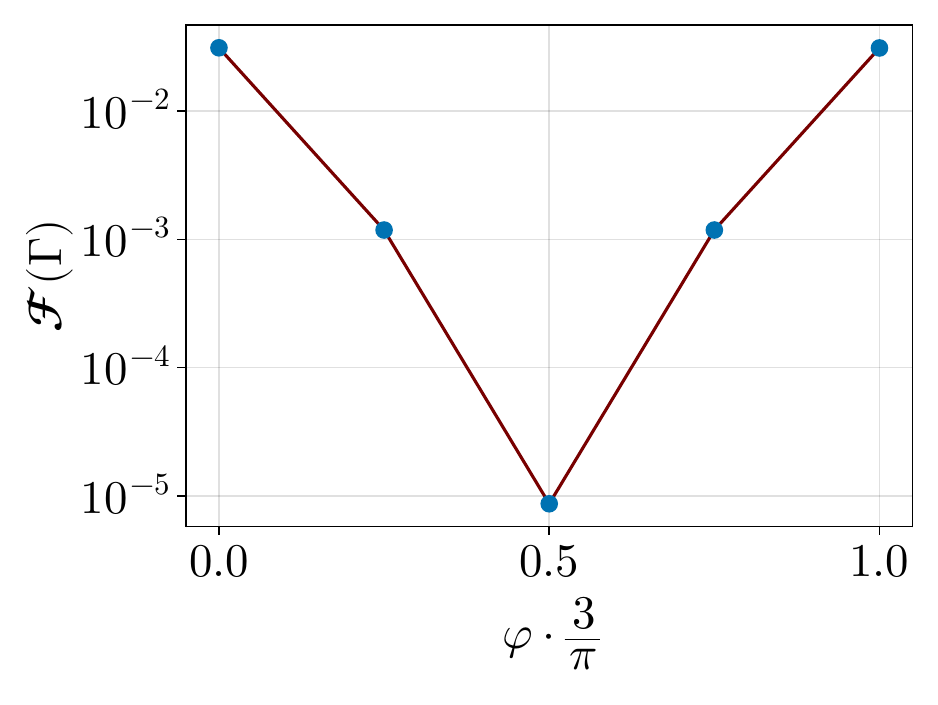}
    \caption{
    Todo
    $\mu/2 = 0.3, L=12$
    }
\end{figure}%
\begin{figure}[htb!]
    \centering
    \includegraphics[width=0.5\linewidth]{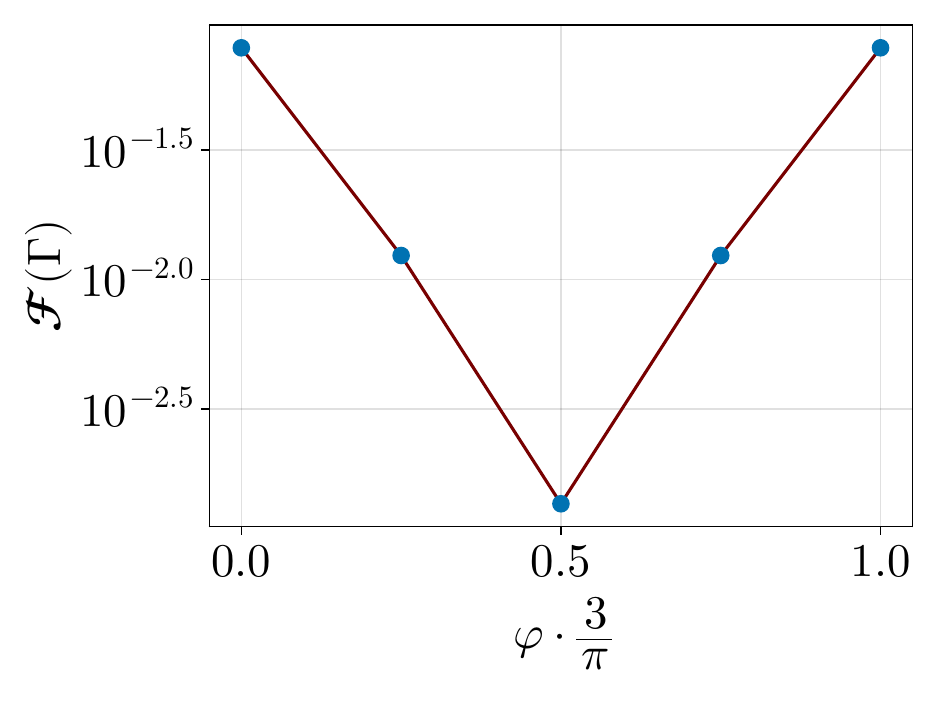}
    \caption{
    Todo
    $\mu/2 = 0.5, L=12$
    }
\end{figure}

\fi

\section{\label{sec:concl}Conclusions}

In this work, we have presented a novel variational ansatz to search for strong and almost strong zero modes in generic Hamiltonians.
Understanding whether a system exhibits such a mode, and what the expected lifetimes are, is of high interest, in particular for realizing topologically protected quantum computers and quantum memories.
Our ansatz is tailored to operate directly in the correct manifold of parafermionic operators by applying a locality-preserving unitary circuit to a local seed operator.
We have demonstrated analytically and numerically that this ansatz can reproduce the correct integrable limits in the Majorana case.
In the extended analysis to generic parameter points of the non-integrable model, we have demonstrated the existence of an edge-localized operator that saturates the commutator with the Hamiltonian to a small but finite value.
This indicates a finite lifetime of edge excitations, which is limited by the value of the commutator.
Furthermore, we have demonstrated for the first time that the chiral Potts model generically hosts ASZMs with finite but prolonged lifetimes in the parameter space associated with a topologically degenerate ground-state manifold.
In addition, we have shown that these ASZMs become true SZMs at special points in the parameter space, as predicted.

In future research, it would be valuable to perform the same analysis for a broader range of parameters.
This could help in understanding how the almost strong zero mode connects to the integrable limits.
Another compelling direction would be to study whether a similar ansatz can be used to search for more general integrals of motion, both in the exact case as with the XYZ model~\cite{Fendley2025} or in approximate ones such as many-body localized systems, similar to~\cite{Pollmann2016, Pancotti2018}.
Finally, the results presented for the non-integrable case display convergence with respect to the block size and number of sites.
However, to obtain more conclusive results, a comprehensive study on convergence is required.
In this context, it would be worthwhile to study different types of circuit structures, such as adding more layers or shifting the unitaries with respect to each other."

\section{Acknowledgments}

We acknowledge fruitful discussions with D. Alcalde, S. Diehl, G. Kells, D. Korbany, A. Rosch, S. Trebst, E. Weerda.
Furthermore, the authors thank Lukas Reich for generating the data for the n=3 parafermionic model.
The simulations presented in this work were produced with a code based on the \rm{ITensor} library~\cite{Fishmann2022}. Data and code are available at~\cite{Tausendpfund2026_dataset}.
The authors acknowledge the support from the DFG under Germany's Excellence Strategy - Cluster of Excellence Matter and Light for Quantum Computing (ML4Q) EXC 2004\slash 1 – 390534769 and project Grant No. 277101999 within the CRC network TR 183.
The authors gratefully acknowledge the Gauss Centre for Supercomputing e.V. (GCS, \href{www.gauss-centre.eu}{www.gauss-centre.eu}) for providing computing time through the John von Neumann Institute for Computing (NIC) on the GCS Supercomputer JUPITER, and the FZJ/VSR for computing time on JURECA~\cite{JURECA2021} at Jülich Supercomputing Centre (JSC) under grant no. SyQMa2TeNs.


\appendix

\section{\label{app:explicit_representations_spin_mat}Spin Representation of Parafermionic Chains}

\if\includeappendixb1

In the spin representation, the $\Z_n$ parafermions $\chi_j$ and $\psi_j$ introduced in Sec.~\ref{sec:parafermion_aszm} are representable by spin operators $\sigma_j$ and $\tau_j$ acting on a $n$ dimensional spin space.
These spin operators are connected to the parafermions by the non-local transformation in Eq.~\eqref{eq:jw_traffo_gen}.
While commuting for $j\neq k$, they fulfill the parafermionic exchange statistics on site with
\begin{equation}\label{eq:parafermions_spin_commutations}
\begin{split}
  \sigma_j^n &= \tau_j^n = \unit\eqcmm\ \sigma_j^\dag = \sigma_j^{(n-1)}\eqcmm\ \tau_j^\dag = \tau_j^{(n-1)}\\
  \sigma_j \tau_j &= \omega \tau_j \sigma_j 
\end{split}
\end{equation}
where $\omega = \exp(2\pi i/n)$.
Owing to this relations, one can chose a basis where either $\sigma_j$ or $\tau_j$ is diagonal.
Choosing $\tau_j$ to be diagonal, the first line Eq.~\eqref{eq:parafermions_spin_commutations} requires the eigenstates of $\tau_j$ to have eigenvalues that are $n$-th roots of unity
\[
\tau_j \ket{l} = \omega^l \ket{l}\eqcmm \ \omega = e^{2\pi i/n}\eqstop
\]
In matrix form, this sets $\tau_j$ to be
\begin{equation}
    \tau_j = \begin{pmatrix}
        1 & 0 & 0 & \dots & 0\\
        0 & \omega & 0 & \dots & 0 \\
        0 & 0 & \omega^2 & \dots & 0 \\
        \vdots & & \ddots & & \vdots \\
        0 & 0 & 0 & \dots & \omega^{n-1}
    \end{pmatrix}
\end{equation}
The second equation of Eq.~\eqref{eq:parafermions_spin_commutations} then requires $\sigma_j$ to act as a shift operator in this basis
\[
\sigma_j \ket{l} = \begin{cases}
    \ket{l-1} \ \text{for}\ l > 0\\
    \ket{n} \ \text{for}\ l = 0
\end{cases}\eqstop
\]
In matrix form, this is given by
\begin{equation}
    \sigma_j = \begin{pmatrix}
        0 & 1 & 0 & \dots & 0\\
        0 & 0 & 1 & \dots & 0\\ 
        \vdots & & \ddots & & \vdots \\
        1 & 0 & 0 & \dots & 0\\
    \end{pmatrix}\eqstop
\end{equation}

In the special case of $n=2$ the algebra simplifies.
First the phase factor becomes $\omega = -1$ and the operators locally anti-commute.
In the local basis where $\tau_j$ is diagonal, the action on the local two-dimensional Hilbert space becomes:
\[
    \tau_j \ket{0} = \ket{0} \,,\ 
    \tau_j \ket{1} = - \ket{1}\eqstop
\]
The off-diagonal operator $\sigma_j$ acts on the local basis as
\[
    \sigma_j\ket{0} = \ket{1}\,,\ \sigma_j\ket{1} = \ket{0}\eqstop
\]
In matrix form, both $\tau_j$ and $\sigma_j$ become
\begin{equation}
    \tau_j = \begin{pmatrix}
        1 & 0 \\
        0 & -1
    \end{pmatrix} \,,\ 
    \sigma_j = \begin{pmatrix}
        0 & 1\\
        1 & 0
    \end{pmatrix}
\end{equation}
and the operators are given by the Pauli-matrices:
\[
\tau_j = \paulz{j}\,,\sigma_j = \paulx{j}\eqstop
\]

\fi

\section{\label{app:derivation_exactly_solvable_limits}Derivation of Integrable Limits}

\if\includeappendixa1

In this appendix, we derive the analytical solution presented in Subsection~\ref{subsec:ex_lim} for the integrable points of the Kitaev-Hubbard chain.
In the spin language, the model is given by 
\begin{equation}\label{eq:kitaev_hubbard_chain_appendix}
    H_{\text{KH}} = \sum_{j=1}^{N-1} -w \paulx{j}\paulx{j+1} + U\paulz{j}\paulz{j+1} - \frac{\mu}{2}\sum_{j=1}^N \paulz{j}\, , 
\end{equation}
see also Eq.~\eqref{eq:kitaev_hubbard_chain}.
%
%
We recall here that even a more general model of interacting Majorana wires, taking the form of a generic XYZ spin chain with arbitrary couplings, is integrable and known to host an exact strong zero mode~\cite{Fendley2016,Fendley2025}. 
However, we will not treat this case here.

The seed operator is set to the Pauli-X operator
\[
\szm_0 = \paulx{1}\eqstop
\]
In both cases, the SZM is representable as an MPO with bond dimension $\chi < 4$; see also Eqs.~\eqref{eq:kitaev_szm} and~\eqref{eq:fendley_szm}.
Thus, there should exists an unitary rotation represented by the quantum circuit
\begin{equation}\label{eq:ciruit_ana}
    \unitary = \tikz[baseline=5.5ex]{
        \tikzset{
            wten/.style={
                draw,
                fill=BurntOrange!20,
                inner sep = 0,
                minimum width = 1.0cm,
                minimum height = 0.6cm
        }}
        \node[wten] (u1) at (0,0) {$u_{1}$};
        \node[wten] (u2) at (0.6,1){$u_{2}$};
        \node[wten] (u3) at (1.2,2){$u_{3}$};
        
        \draw[thick] ($(u1.south) + (-0.3,-0.4)$) -- ($(u1.south) + (-0.3,0)$);
        \draw[thick] ($(u1.south) + ( 0.3,-0.4)$) -- ($(u1.south) + (0.3,0)$);
        \draw[thick] ($(u1.north) + (-0.3, 2.4)$) -- ($(u1.north) + (-0.3,0)$);

        \draw[thick] ($(u1.north) + ( 0.3, 0.0)$) -- ($(u2.south) + (-0.3,0)$);
        \draw[thick] ($(u2.south) + ( 0.3,-1.4)$) -- ($(u2.south) + (0.3,0)$);
        \draw[thick] ($(u2.north) + (-0.3, 1.4)$) -- ($(u2.north) + (-0.3,0)$);

        \draw[thick] ($(u2.north) + ( 0.3, 0.0)$) -- ($(u3.south) + (-0.3,0)$);
        \draw[thick] ($(u3.south) + ( 0.3,-2.4)$) -- ($(u3.south) + (0.3,0)$);
        \draw[thick] ($(u3.north) + (-0.3, 0.4)$) -- ($(u3.north) + (-0.3,0)$);
        \draw[thick] ($(u3.north) + ( 0.3, 0.4)$) -- ($(u3.north) + ( 0.3,0)$);
        \node[anchor=east] at (2.5,1) {$\dots$};
    }      = u_{N-1} u_{N-2}\dots u_{2} u_{1}\eqcmm
\end{equation}
such that $\szm = \unitary\szm_0 \unitary^\dag$ for both cases.
As discussed in Section~\ref{sec:var_ansatz}, the unitaries $u_j$ are chosen to commute with the parity operator $\paraparity = \prod_j \paulz{j}$.
This requires that the total sum of the Pauli-X and Pauli-Y operators is conserved when acting on a pair of Pauli operators.
In particular, by representing $u_j$ as the matrix exponential of an Hermitian operator $V_j$ through
\[
    u_{j} = e^{iV_{j}}\eqcmm
\]
this operator can only be of the form
\[
\begin{split}
V_{j} &= \alpha_0 \unit + \alpha_1 \paulz{j} + \alpha_2 \paulz{j+1} + \alpha_3 \paulz{j}\paulz{j+1}\\
{}&+ \alpha_4 \paulx{j} \paulx{j+1}+ \alpha_5 \pauly{j} \pauly{j+1}
+ \alpha_6 \paulx{j} \pauly{j+1}
+ \alpha_7 \pauly{j} \paulx{j+1}\eqcmm
\end{split}
\]
with $\alpha_j \in \mathds{R}$.

\subsection{Case \texorpdfstring{$U=0$}{U0}}

For $U=0$, Eq.~\eqref{eq:kitaev_hubbard_chain_appendix} reduces to the transverse-field Ising model, and the SZM is given by
\begin{equation}\label{eq:caseI_szm}
    \begin{split}
        \szm &= \mathcal{N}(N) \sum_{j=1}^N \gamma^{j-1}\left(\prod_{k<j} \paulz{k}\right) \paulx{j}\eqcmm\\
        \mathcal{N}(N) &= \sqrt{\frac{1 -
        \gamma^2}{1-\gamma^{2N}}} \eqcmm
    \end{split}
\end{equation}
where $\gamma = \mu/(2w)$.
Since this operator is a sum over strings of the type $(\prod_{k<j} \paulz{k})\paulx{j}$, the two-body unitary should encode the transformation
\[
    \paulx{j}\xrightarrow{u^{\mathrm{I}}_{j,j+1}} \ \xi_1\, \paulx{j} + \xi_2\, \paulz{j}\paulx{j+1}\eqcmm\quad \xi_1^2 + \xi_2^2 = 1\eqstop
\]
In this way, the Pauli-X operator is shifted to the right owing to the ladder structure of the quantum circuit, generating the desired string structure.
A suitable operator that generates this transformation is given by
\begin{equation}\label{eq:unitary}
    u_{j} = e^{i\alpha_j \pauly{j} \paulx{j+1}} = \cos(\alpha_j) \unit + i\sin(\alpha_j) \pauly{j}\paulx{j+1}
\end{equation}
with $\xi_1 = \cos(2\alpha_j)$ and $\xi_2 = \sin(2\alpha_j)$.
An application of the full unitary from Eq.~\eqref{eq:ciruit_ana} on the seed operator $\szm_0=\paulx{1}$ leads to 
\begin{equation}\label{eq:application_circuit}
    \begin{split}
        \unitary\szm_0 \unitary^\dag &=
        \sum_{j=1}^{N-1} \cos(2\alpha_j) \left(\prod_{k=1}^{j-1} \sin(2\alpha_k)\paulz{k}\right)\paulx{j} \\
        {}&+ \left(\prod_{k=1}^{N-1}\sin(2\alpha_k)\paulz{k}\right)\paulx{N}\eqstop
    \end{split}
\end{equation}
The rotation angles $\alpha_j$ are obtained by comparing this to the exact expression of the SZM in Eq.~\eqref{eq:caseI_szm} 
\begin{equation}\label{eq:defining_equation_angle_case_I}
   \begin{split}
        \beta_j &\coloneqq \cos(2\alpha_j) \prod_{k=1}^{j-1} \sin(2\alpha_k) = \mathcal{N}(N)\gamma^{j-1} \\
       {}&= \sqrt{\frac{1 - \gamma^2}{1-\gamma^{2N}}}\gamma^{j-1}\quad\forall\quad 0<j<N\\
        \beta_N &\coloneqq \prod_{k=1}^{N-1} \sin(2\alpha_k) = \mathcal{N}(N)\gamma^{N-1}\\
        {}&= \sqrt{\frac{1 - \gamma^2}{1-\gamma^{2N}}}\gamma^{N-1}\eqstop
    \end{split}
\end{equation}
Dividing $\beta_N$ by $\beta_{N-1}$ results in
\[
   \frac{\beta_{N}}{\beta_{N-1}} = \tan(2\alpha_{N-1}) = \gamma\eqcmm
\]
fixing the last rotation angle $\alpha_{N-1}$.
To find the result for the remaining $N-2$ angles, consider the relation
\[
   \sum_{l=0}^r\beta_{N-l}^2 = \prod_{k = 1}^{N-r-1} \sin(2\alpha_k)^2 = \mathcal{N}(N)^2 \sum_{l=0}^{r} \gamma^{2(N - l - 1)}\eqcmm
\]
which can be proven by induction.
Dividing this sum by $\beta_{N-r-1}^2$ results in
\[
   \frac{\sum_{l=0}^r\beta_{N-l}^2}{\beta_{N-r-1}^2} = \tan(2\alpha_{N-r-1})^2 = \frac{\sum_{l=0}^{r} \gamma^{2(N - l - 1)}}{\gamma^{2(N-r-2)}}\eqstop
\]
Finally, setting $j = N-r-1$ gives the solution for all angles for $1\le j\le N-2$
\[
   \tan(2\alpha_j)^2 = \gamma^2 \sum_{l=0}^{N-j-1}\gamma^{2(N - j -1 - l)} = \gamma^2 \sum_{k=0}^{N-j-1} \gamma^{2k}\eqstop
\]
The last step follows after a simple change in the summation variable $k = N-j-1-l$.
Putting everything together, the solution for all angles can be written as
\begin{equation}\label{eq:angle_solution}
    \begin{split}
        \tan(2\alpha_j)^2 &= \rho^2\sum_{k=0}^{N-j-1} \gamma^{2k}\\
        {} &= \frac{\gamma^2}{1- \gamma^2}\left(1 - \gamma^{2(N-j)}\right)\eqstop
    \end{split}
\end{equation}

\subsection{Case \texorpdfstring{$\mu=0$}{mu0}}

In the limit $\mu = 0$, the Kitaev-Hubbard chain of Eq.~\eqref{eq:kitaev_hubbard_chain_appendix} possess a SZM given by~\cite{Fendley2016}
\begin{equation}\label{eq:caseII_szm}
\begin{split}
    \szm &= \mathcal{N}(N) \sum_{j=1}^{\left\lfloor\frac{N + 1}{2}\right\rfloor} (-\gamma)^{j-1}\left(\prod_{k<j} \pauly{2k-1}\pauly{2k}\right) \paulx{2j-1}\eqcmm\\
    \mathcal{N}(N) &= \sqrt{\frac{1 -
    \gamma^2}{1-\gamma^{2\left\lfloor\frac{N + 1}{2}\right\rfloor}}} \eqcmm
\end{split}
\end{equation}
with $\gamma = U/w$.
Similar to the previous case of $U=0$, the operator $\szm$ is a sum over strings of the form $(\prod_{k<l} \pauly{k})\paulx{l}$.
However, the sum only runs over all odd indices $l = 2j-1$ of the chain.
To obtain the correct unitary rotation in this case, it is necessary to consider the combined action of two adjacent unitary matrices of our ansatz from Eq.~\eqref{eq:unitary_ansatz} acting on the pairs of indices $(2j-1,2j)$ and $(2j,2j+1)$.
A possible chain of parity conserving transformations is given by
\begin{equation}\label{eq:transformation_chain}
    \begin{split}
    \paulx{2j-1}&\xrightarrow{u_{2j-1}} \ \xi_1\, \paulx{2j-1} + \xi_2\, \pauly{2j-1}\paulz{2j}\\
    {}&\xrightarrow{u_{2j}}\xi_1\, \paulx{2j-1} + \xi_2\, \pauly{2j-1}\pauly{2j}\paulx{2j+1} \eqcmm
    \end{split}
\end{equation}
with the constrain $\xi_1^2 + \xi_2^2 = 1$.
If this chain of transformation exists, the ladder structure ensures that the Pauli-X operator is shifted to the right by two lattice sites while generating the desired string structure, similar to the case of $U=0$.

First, consider the transformation
\[
\paulx{2j-1}\xrightarrow{u_{2j-1}} \ \xi_1\, \paulx{2j-1} + \xi_2\pauly{2j-1}\paulz{2j}\eqstop
\]
A unitary that performs this rotation of Pauli matrices is given by
\begin{equation}\label{eq:odd_rotation}
    \begin{split}
    u_{2j-1} &= e^{i\alpha_j \paulz{2j-1} \paulz{2j}}\\
    {}&= \cos(\alpha_j) \unit + i\sin(\alpha_j) \paulz{2j-1}\paulz{2j}
    \end{split}
\end{equation}
with $\xi_1 = \cos(2\alpha_j)$ and $\xi_2 = -\sin(2\alpha_j)$.
The second transformation acting as
\[
\paulz{2j}\xrightarrow{u_{2j}}\pauly{2j}\paulx{2j+1}\eqcmm
\]
is generated by the unitary matrix 
\begin{equation}\label{eq:even_rotation}
    u_{2j} = e^{i\frac{\pi}{4} \paulx{2j}\paulx{2j+1}} = 
    \frac{1}{\sqrt{2}}\left(
        \unit + i\paulx{2j}\paulx{2j+1}
    \right)\eqcmm
\end{equation}
where no additional parameters are required.
Combining the two unitary matrices from Eqs.~\eqref{eq:odd_rotation} and~\eqref{eq:even_rotation}, the transformation chain from Eq.~\eqref{eq:transformation_chain} is obtained.

Assume that $N = 2\tilde{N} + 1$ is odd~\footnote{The case of an even number of sites can be treated on equal footing, but requires a slightly modified two-body unitary transformation on the last two sites.}.
In this case, the full unitary circuit $\unitary$ in Eq.~\eqref{eq:ciruit_ana} involves an even number of two-body unitaries.
Applying the full unitary circuit to the seed operator $\szm_0=\paulx{1}$ leads to
\begin{equation}\label{eq:application_circuit_case2}
\begin{split}
    \unitary\szm_0 \unitary^\dag =
    \sum_{j=1}^{\tilde{N}} &(-1)^{j-1} \cos(2\alpha_j) \\
    {}&\left(\prod_{k=1}^{j-1} \sin(2\alpha_k)\pauly{2k-1}\pauly{2k}\right)\paulx{2j-1} \\
    {}+ (-1)^{\tilde{N}}&\left(\prod_{k=1}^{\tilde{N}}\sin(2\alpha_k)\pauly{2k-1}\pauly{2k}\right)\paulx{2\tilde{N}+1}\eqstop
\end{split}
\end{equation}
Again, the unknown rotation angles are fixed by comparing this to the exact expression of the SZM in Eq.~\eqref{eq:caseII_szm}
\begin{equation}\label{eq:defining_equation_angle_case_II_1}
   \begin{split}
        \beta_j &= \cos(2\alpha_j) \prod_{k=1}^{j-1} \sin(2\alpha_k) = \mathcal{N}(N)\gamma^{j-1} \\
        {}&= \sqrt{\frac{1 - \gamma^2}{1-\gamma^{2(\tilde{N} + 1)}}}\gamma^{j-1}\quad\forall\quad 0<j\le\tilde{N}\\
    \end{split}
\end{equation}
and
\begin{equation}\label{eq:defining_equation_angle_case_II_2}
   \begin{split}
        \beta_{\tilde{N} + 1} &= \prod_{k=1}^{\tilde{N}} \sin(2\alpha_k) \\
        {}&= \mathcal{N}(N)\gamma^{\tilde{N}} = \sqrt{\frac{1 - \gamma^2}{1-\gamma^{2(\tilde{N} + 1)}}}\gamma^{\tilde{N}}\eqstop
    \end{split}
\end{equation}
This set of equations is similar to the set of equations~\eqref{eq:defining_equation_angle_case_I} found in the case $U=0$, and can be solved in an analogous manner.
The solution for the angles is given by for all $1\le j \le \tilde{N}$
\begin{equation}\label{eq:angle_solution_caseII}
\begin{split}
    \tan(2\alpha_j)^2 &= \gamma^2\sum_{k=0}^{\tilde{N}-j} \gamma^{2k} \\
    {}&= \frac{\gamma^2}{1- \gamma^2}\left(1 - \gamma^{2(\tilde{N}+1-j)}\right)\eqstop
\end{split}
\end{equation}

Equations~\eqref{eq:angle_solution} and~\eqref{eq:angle_solution_caseII} conclude the derivation of the analytical solution for the SZM presented in the main text in Subsection~\ref{subsec:ex_lim}.

\fi

\bibliography{refs.bib}

\end{document}

%% file: auxiliaries/package_definition.tex
\usepackage[english]{babel}
\usepackage{lmodern}

\usepackage[dvipsnames]{xcolor}

\usepackage{amsmath, amssymb}

\usepackage{graphicx}
\usepackage[colorlinks=true, allcolors=Maroon]{hyperref}
\usepackage[utf8]{inputenc}
\usepackage{dsfont}
\usepackage[T1]{fontenc}
\usepackage{braket}
\usepackage{mathtools}

\usepackage{algorithm}
\usepackage{algpseudocode}

\usepackage{enumitem}

\usepackage{tabularx}


%% file: auxiliaries/tikz_aux.tex
\usepackage{tikz}
\usepackage{pgfplots}
\usepackage{pgfplotstable}

\usetikzlibrary{decorations.pathmorphing}
\usetikzlibrary{decorations.pathreplacing,calligraphy}
\usetikzlibrary{decorations.markings}
\usetikzlibrary{calc,fit,patterns}

\pgfplotsset{compat=1.18}
\pgfplotsset{every axis/.append style={
    label style = {font = \large},
    scale = 1,
    grid = major,
    grid style = dashed,
    mark size = 2pt,
}}

\pgfplotsset{every axis plot/.append style = {
    mark=*, 
    only marks,
}}

\tikzstyle{unit}=[circle,draw=red!50,fill=red!20,thick]
\tikzstyle{tensor}=[rectangle,draw=blue!50,fill=blue!20,thick, minimum width=6mm, minimum height=6mm]
\tikzstyle{newTensor}=[rectangle,draw=magenta!50,fill=magenta!20,thick, minimum width=6mm, minimum height=6mm]
\tikzstyle{nTensor}=[rectangle,draw=blue!50,fill=blue!20,thick,minimum width=20mm,rounded corners=.8ex]
\tikzstyle{colorFill}=[-,fill=magenta!20]
\tikzstyle{trotterTensor} = [rectangle,draw=red!50,fill=red!20,thick, minimum width=16mm, minimum height=6mm]

\tikzstyle{graphnode} = [shape=circle,draw,inner sep=2pt]

\tikzstyle{graphedge} = [->, >=stealth]

\tikzset{
    aten/.style={
        draw,
        fill=MidnightBlue!20,
        inner sep = 0,
        minimum width = 0.6cm,
        minimum height = 0.6cm
}}

\tikzset{
    bten/.style={
        draw,
        fill=OliveGreen!20,
        inner sep = 0,
        minimum width = 0.6cm,
        minimum height = 0.6cm
}}
\tikzset{
    cten/.style={
        draw,
        fill=Dandelion!20,
        inner sep = 0,
        minimum width = 0.6cm,
        minimum height = 0.6cm
}}
\tikzset{
    dten/.style={
        draw,
        fill=Violet!20,
        inner sep = 0,
        minimum width = 0.6cm,
        minimum height = 0.6cm
}}
\tikzset{
    diagten/.style={
        draw,
        fill=Red!20,
        inner sep = 0,
        minimum width = 0.6cm,
        minimum height = 0.6cm
}}


%% file: auxiliaries/additional_declarations.tex
\DeclareMathOperator{\Z}{\mathds{Z}}

\newcommand{\vdag}{{\phantom\dag}}
\newcommand{\hc}{\mathrm{h.c.}}

\newcommand{\paulx}[1]{\sigma^x_{#1}}
\newcommand{\pauly}[1]{\sigma^y_{#1}}
\newcommand{\paulz}[1]{\sigma^z_{#1}}

\DeclareMathOperator{\Tr}{\mathrm{Tr}}

\newcommand{\hilb}{\mathcal{H}}
\newcommand{\liouv}{\mathcal{L}}
\newcommand{\paraparity}{\hat{\omega}^P}

\DeclareMathOperator{\unit}{\mathds{1}}

\DeclareMathOperator{\szm}{\Gamma}
\DeclareMathOperator{\paraman}{\mathcal{M}}
\DeclareMathOperator{\costfn}{\mathcal{F}}
\DeclareMathOperator{\unitary}{\mathcal{U}}
\DeclareMathOperator{\chiralangle}{\varphi}

\newcommand{\eqstop}{\, .}
\newcommand{\eqcmm}{\, ,}

%% file: refs.bib
@article{Kitaev2001,
doi = {10.1070/1063-7869/44/10S/S29},
url = {https://dx.doi.org/10.1070/1063-7869/44/10S/S29},
year = {2001},
month = {oct},
publisher = {},
volume = {44},
number = {10S},
pages = {131},
author = {A Yu Kitaev},
title = {Unpaired Majorana fermions in quantum
wires},
journal = {Physics-Uspekhi}
}

@article{Nayak2008,
  title = {Non-Abelian anyons and topological quantum computation},
  author = {Nayak, Chetan and Simon, Steven H. and Stern, Ady and Freedman, Michael and Das Sarma, Sankar},
  journal = {Rev. Mod. Phys.},
  volume = {80},
  issue = {3},
  pages = {1083--1159},
  numpages = {0},
  year = {2008},
  month = {Sep},
  publisher = {American Physical Society},
  doi = {10.1103/RevModPhys.80.1083},
  url = {https://link.aps.org/doi/10.1103/RevModPhys.80.1083}
}

@article{Ivanov2001,
  title = {Non-Abelian Statistics of Half-Quantum Vortices in $\mathit{p}$-Wave Superconductors},
  author = {Ivanov, D. A.},
  journal = {Phys. Rev. Lett.},
  volume = {86},
  issue = {2},
  pages = {268--271},
  numpages = {0},
  year = {2001},
  month = {Jan},
  publisher = {American Physical Society},
  doi = {10.1103/PhysRevLett.86.268},
  url = {https://link.aps.org/doi/10.1103/PhysRevLett.86.268}
}

@article{Alicea2011,
	author = {Alicea, Jason and Oreg, Yuval and Refael, Gil and von Oppen, Felix and Fisher, Matthew P. A.},
	title = {{Non-Abelian statistics and topological quantum information processing in 1D wire networks}},
	journal = {Nat. Phys.},
	volume = {7},
	number = {5},
	pages = {412--417},
	year = {2011},
	month = may,
	issn = {1745-2481},
	publisher = {Nature Publishing Group},
	doi = {10.1038/nphys1915}
}

@article{Sarma2015,
	author = {Sarma, Sankar Das and Freedman, Michael and Nayak, Chetan},
	title = {{Majorana zero modes and topological quantum computation}},
	journal = {npj Quantum Inf.},
	volume = {1},
	number = {15001},
	pages = {1--13},
	year = {2015},
	month = oct,
	issn = {2056-6387},
	publisher = {Nature Publishing Group},
	doi = {10.1038/npjqi.2015.1}
}

@article{Turner2011,
  title = {Topological phases of one-dimensional fermions: An entanglement point of view},
  author = {Turner, Ari M. and Pollmann, Frank and Berg, Erez},
  journal = {Phys. Rev. B},
  volume = {83},
  issue = {7},
  pages = {075102},
  numpages = {11},
  year = {2011},
  month = {Feb},
  publisher = {American Physical Society},
  doi = {10.1103/PhysRevB.83.075102},
  url = {https://link.aps.org/doi/10.1103/PhysRevB.83.075102}
}

@article{Ryu2010,
	author = {Ryu, Shinsei and Schnyder, Andreas P. and Furusaki, Akira and Ludwig, Andreas W. W.},
	title = {{Topological insulators and superconductors: tenfold way and dimensional hierarchy}},
	journal = {New J. Phys.},
	volume = {12},
	number = {6},
	pages = {065010},
	year = {2010},
	month = jun,
	issn = {1367-2630},
	publisher = {IOP Publishing},
	doi = {10.1088/1367-2630/12/6/065010}
}

@article{Altland1997,
  title = {Nonstandard symmetry classes in mesoscopic normal-superconducting hybrid structures},
  author = {Altland, Alexander and Zirnbauer, Martin R.},
  journal = {Phys. Rev. B},
  volume = {55},
  issue = {2},
  pages = {1142--1161},
  numpages = {0},
  year = {1997},
  month = {Jan},
  publisher = {American Physical Society},
  doi = {10.1103/PhysRevB.55.1142},
  url = {https://link.aps.org/doi/10.1103/PhysRevB.55.1142}
}

@article{Sen13,
  title = {Floquet generation of Majorana end modes and topological invariants},
  author = {Thakurathi, Manisha and Patel, Aavishkar A. and Sen, Diptiman and Dutta, Amit},
  journal = {Phys. Rev. B},
  volume = {88},
  issue = {15},
  pages = {155133},
  numpages = {13},
  year = {2013},
  month = {Oct},
  publisher = {American Physical Society},
  doi = {10.1103/PhysRevB.88.155133},
  url = {https://link.aps.org/doi/10.1103/PhysRevB.88.155133}
}

@article{Fendley2016,
   author = {Paul Fendley},
   doi = {10.1088/1751-8113/49/30/30LT01},
   issn = {17518121},
   issue = {30},
   journal = {Journal of Physics A: Mathematical and Theoretical},
   month = {6},
   publisher = {Institute of Physics Publishing},
   title = {Strong zero modes and eigenstate phase transitions in the XYZ/interacting Majorana chain},
   volume = {49},
   year = {2016},
}

@article{Fendley2012,
doi = {10.1088/1742-5468/2012/11/P11020},
url = {https://dx.doi.org/10.1088/1742-5468/2012/11/P11020},
year = {2012},
month = {nov},
publisher = {IOP Publishing and SISSA},
volume = {2012},
number = {11},
pages = {P11020},
author = {Paul Fendley},
title = {Parafermionic edge zero modes in Zn-invariant spin chains},
journal = {Journal of Statistical Mechanics: Theory and Experiment}
}

@article{Yates2019,
  title = {Almost strong ($0,\ensuremath{\pi}$) edge modes in clean interacting one-dimensional Floquet systems},
  author = {Yates, Daniel J. and Essler, Fabian H. L. and Mitra, Aditi},
  journal = {Phys. Rev. B},
  volume = {99},
  issue = {20},
  pages = {205419},
  numpages = {13},
  year = {2019},
  month = {May},
  publisher = {American Physical Society},
  doi = {10.1103/PhysRevB.99.205419},
  url = {https://link.aps.org/doi/10.1103/PhysRevB.99.205419}
}

@article{Vernier24,
  title = {Strong Zero Modes in Integrable Quantum Circuits},
  author = {Vernier, Eric and Yeh, Hsiu-Chung and Piroli, Lorenzo and Mitra, Aditi},
  journal = {Phys. Rev. Lett.},
  volume = {133},
  issue = {5},
  pages = {050606},
  numpages = {7},
  year = {2024},
  month = {Aug},
  publisher = {American Physical Society},
  doi = {10.1103/PhysRevLett.133.050606},
  url = {https://link.aps.org/doi/10.1103/PhysRevLett.133.050606}
}

@article{Else2017,
  title = {Prethermal Strong Zero Modes and Topological Qubits},
  author = {Else, Dominic V. and Fendley, Paul and Kemp, Jack and Nayak, Chetan},
  journal = {Phys. Rev. X},
  volume = {7},
  issue = {4},
  pages = {041062},
  numpages = {22},
  year = {2017},
  month = {Dec},
  publisher = {American Physical Society},
  doi = {10.1103/PhysRevX.7.041062},
  url = {https://link.aps.org/doi/10.1103/PhysRevX.7.041062}
}

@article{Parker2019,
  title = {Topologically Protected Long Edge Coherence Times in Symmetry-Broken Phases},
  author = {Parker, Daniel E. and Vasseur, Romain and Scaffidi, Thomas},
  journal = {Phys. Rev. Lett.},
  volume = {122},
  issue = {24},
  pages = {240605},
  numpages = {5},
  year = {2019},
  month = {Jun},
  publisher = {American Physical Society},
  doi = {10.1103/PhysRevLett.122.240605},
  url = {https://link.aps.org/doi/10.1103/PhysRevLett.122.240605}
}

@article{Kemp2020,
  title = {Symmetry-Enhanced Boundary Qubits at Infinite Temperature},
  author = {Kemp, Jack and Yao, Norman Y. and Laumann, Chris R.},
  journal = {Phys. Rev. Lett.},
  volume = {125},
  issue = {20},
  pages = {200506},
  numpages = {6},
  year = {2020},
  month = {Nov},
  publisher = {American Physical Society},
  doi = {10.1103/PhysRevLett.125.200506},
  url = {https://link.aps.org/doi/10.1103/PhysRevLett.125.200506}
}

@article{Yates2020,
  title = {Lifetime of Almost Strong Edge-Mode Operators in One-Dimensional, Interacting, Symmetry Protected Topological Phases},
  author = {Yates, Daniel J. and Abanov, Alexander G. and Mitra, Aditi},
  journal = {Phys. Rev. Lett.},
  volume = {124},
  issue = {20},
  pages = {206803},
  numpages = {6},
  year = {2020},
  month = {May},
  publisher = {American Physical Society},
  doi = {10.1103/PhysRevLett.124.206803},
  url = {https://link.aps.org/doi/10.1103/PhysRevLett.124.206803}
}

@article{Yates2020_B,
  title = {Dynamics of almost strong edge modes in spin chains away from integrability},
  author = {Yates, Daniel J. and Abanov, Alexander G. and Mitra, Aditi},
  journal = {Phys. Rev. B},
  volume = {102},
  issue = {19},
  pages = {195419},
  numpages = {15},
  year = {2020},
  month = {Nov},
  publisher = {American Physical Society},
  doi = {10.1103/PhysRevB.102.195419},
  url = {https://link.aps.org/doi/10.1103/PhysRevB.102.195419}
}

@article{Yeh2023,
  title = {Slowly decaying zero mode in a weakly nonintegrable boundary impurity model},
  author = {Yeh, Hsiu-Chung and Cardoso, Gabriel and Korneev, Leonid and Sels, Dries and Abanov, Alexander G. and Mitra, Aditi},
  journal = {Phys. Rev. B},
  volume = {108},
  issue = {16},
  pages = {165143},
  numpages = {16},
  year = {2023},
  month = {Oct},
  publisher = {American Physical Society},
  doi = {10.1103/PhysRevB.108.165143},
  url = {https://link.aps.org/doi/10.1103/PhysRevB.108.165143}
}

@article{Olund2025,
  title = {Boundary strong zero modes},
  author = {Olund, Christopher T. and Yao, Norman Y. and Kemp, Jack},
  journal = {Phys. Rev. B},
  volume = {111},
  issue = {20},
  pages = {L201114},
  numpages = {6},
  year = {2025},
  month = {May},
  publisher = {American Physical Society},
  doi = {10.1103/PhysRevB.111.L201114},
  url = {https://link.aps.org/doi/10.1103/PhysRevB.111.L201114}
}

@article{Tausendpfund2025,
  title = {Almost strong zero modes at finite temperature},
  author = {Tausendpfund, Niklas and Mitra, Aditi and Rizzi, Matteo},
  journal = {Phys. Rev. Res.},
  volume = {7},
  issue = {2},
  pages = {023245},
  numpages = {19},
  year = {2025},
  month = {Jun},
  publisher = {American Physical Society},
  doi = {10.1103/PhysRevResearch.7.023245},
  url = {https://link.aps.org/doi/10.1103/PhysRevResearch.7.023245}
}

@phdthesis{Tausendpfund2025_B,
        year = {2025},
      school = {Universit{\"a}t zu K{\"o}ln},
      author = {Niklas Tausendpfund},
       title = {On the Design and Characterization of Exotic Phases in One Dimension},
    url = {https://kups.ub.uni-koeln.de/79055/}
}

@article{Alicea2016,
   author = "Alicea, Jason and Fendley, Paul",
   title = "Topological Phases with Parafermions: Theory and Blueprints", 
   journal= "Annual Review of Condensed Matter Physics",
   year = "2016",
   volume = "7",
   number = "Volume 7, 2016",
   pages = "119-139",
   doi = "https://doi.org/10.1146/annurev-conmatphys-031115-011336",
   url = "https://www.annualreviews.org/content/journals/10.1146/annurev-conmatphys-031115-011336",
   publisher = "Annual Reviews",
   issn = "1947-5462",
   type = "Journal Article",
  }

@article{Iemini2017_B,
  title = {Topological Phases of Parafermions: A Model with Exactly Solvable Ground States},
  author = {Iemini, Fernando and Mora, Christophe and Mazza, Leonardo},
  journal = {Phys. Rev. Lett.},
  volume = {118},
  issue = {17},
  pages = {170402},
  numpages = {6},
  year = {2017},
  month = {Apr},
  publisher = {American Physical Society},
  doi = {10.1103/PhysRevLett.118.170402},
  url = {https://link.aps.org/doi/10.1103/PhysRevLett.118.170402}
}

@article{Kells2018,
  title = {Localization enhanced and degraded topological order in interacting $p$-wave wires},
  author = {Kells, G. and Moran, N. and Meidan, D.},
  journal = {Phys. Rev. B},
  volume = {97},
  issue = {8},
  pages = {085425},
  numpages = {15},
  year = {2018},
  month = {Feb},
  publisher = {American Physical Society},
  doi = {10.1103/PhysRevB.97.085425},
  url = {https://link.aps.org/doi/10.1103/PhysRevB.97.085425}
}

@article{Stoudenmire2011,
  title = {Interaction effects in topological superconducting wires supporting Majorana fermions},
  author = {Stoudenmire, E. M. and Alicea, Jason and Starykh, Oleg A. and Fisher, Matthew P.A.},
  journal = {Phys. Rev. B},
  volume = {84},
  issue = {1},
  pages = {014503},
  numpages = {13},
  year = {2011},
  month = {Jul},
  publisher = {American Physical Society},
  doi = {10.1103/PhysRevB.84.014503},
  url = {https://link.aps.org/doi/10.1103/PhysRevB.84.014503}
}

@article{Katsura2015,
  title = {Exact ground states and topological order in interacting Kitaev/Majorana chains},
  author = {Katsura, Hosho and Schuricht, Dirk and Takahashi, Masahiro},
  journal = {Phys. Rev. B},
  volume = {92},
  issue = {11},
  pages = {115137},
  numpages = {12},
  year = {2015},
  month = {Sep},
  publisher = {American Physical Society},
  doi = {10.1103/PhysRevB.92.115137},
  url = {https://link.aps.org/doi/10.1103/PhysRevB.92.115137}
}

@article{Calzona2018,
  title = {${\mathbb{Z}}_{4}$ parafermions in one-dimensional fermionic lattices},
  author = {Calzona, Alessio and Meng, Tobias and Sassetti, Maura and Schmidt, Thomas L.},
  journal = {Phys. Rev. B},
  volume = {98},
  issue = {20},
  pages = {201110},
  numpages = {6},
  year = {2018},
  month = {Nov},
  publisher = {American Physical Society},
  doi = {10.1103/PhysRevB.98.201110},
  url = {https://link.aps.org/doi/10.1103/PhysRevB.98.201110}
}

@article{Chew2018,
  title = {Fermionized parafermions and symmetry-enriched Majorana modes},
  author = {Chew, Aaron and Mross, David F. and Alicea, Jason},
  journal = {Phys. Rev. B},
  volume = {98},
  issue = {8},
  pages = {085143},
  numpages = {36},
  year = {2018},
  month = {Aug},
  publisher = {American Physical Society},
  doi = {10.1103/PhysRevB.98.085143},
  url = {https://link.aps.org/doi/10.1103/PhysRevB.98.085143}
}

@article{Jermyn2014,
  title = {Stability of zero modes in parafermion chains},
  author = {Jermyn, Adam S. and Mong, Roger S. K. and Alicea, Jason and Fendley, Paul},
  journal = {Phys. Rev. B},
  volume = {90},
  issue = {16},
  pages = {165106},
  numpages = {14},
  year = {2014},
  month = {Oct},
  publisher = {American Physical Society},
  doi = {10.1103/PhysRevB.90.165106},
  url = {https://link.aps.org/doi/10.1103/PhysRevB.90.165106}
}

@article{White1992,
  title = {Density matrix formulation for quantum renormalization groups},
  author = {White, Steven R.},
  journal = {Phys. Rev. Lett.},
  volume = {69},
  issue = {19},
  pages = {2863--2866},
  numpages = {0},
  year = {1992},
  month = {Nov},
  publisher = {American Physical Society},
  doi = {10.1103/PhysRevLett.69.2863},
  url = {https://link.aps.org/doi/10.1103/PhysRevLett.69.2863}
}

@article{Cirac2017,
doi = {10.1088/1742-5468/aa7e55},
url = {https://doi.org/10.1088/1742-5468/aa7e55},
year = {2017},
month = {aug},
publisher = {IOP Publishing and SISSA},
volume = {2017},
number = {8},
pages = {083105},
author = {Ignacio Cirac, J and Perez-Garcia, David and Schuch, Norbert and Verstraete, Frank},
title = {Matrix product unitaries: structure, symmetries, and topological invariants},
journal = {Journal of Statistical Mechanics: Theory and Experiment}
}

@article{Shukla2025,
    author = {Shukla, Sujeet K.},
    title = {A simple and general equation for matrix product unitary generation},
    journal = {Journal of Mathematical Physics},
    volume = {66},
    number = {10},
    pages = {101901},
    year = {2025},
    month = {10},
    issn = {0022-2488},
    doi = {10.1063/5.0262317},
    url = {https://doi.org/10.1063/5.0262317}
}

@article{Wahl2017,
  title = {Efficient Representation of Fully Many-Body Localized Systems Using Tensor Networks},
  author = {Wahl, Thorsten B. and Pal, Arijeet and Simon, Steven H.},
  journal = {Phys. Rev. X},
  volume = {7},
  issue = {2},
  pages = {021018},
  numpages = {17},
  year = {2017},
  month = {May},
  publisher = {American Physical Society},
  doi = {10.1103/PhysRevX.7.021018},
  url = {https://link.aps.org/doi/10.1103/PhysRevX.7.021018}
}

@article{Pollmann2016,
  title = {Efficient variational diagonalization of fully many-body localized Hamiltonians},
  author = {Pollmann, Frank and Khemani, Vedika and Cirac, J. Ignacio and Sondhi, S. L.},
  journal = {Phys. Rev. B},
  volume = {94},
  issue = {4},
  pages = {041116},
  numpages = {5},
  year = {2016},
  month = {Jul},
  publisher = {American Physical Society},
  doi = {10.1103/PhysRevB.94.041116},
  url = {https://link.aps.org/doi/10.1103/PhysRevB.94.041116}
}

@Article{Hauru2021,
	title={{Riemannian optimization of isometric tensor networks}},
	author={Markus Hauru and Maarten Van Damme and Jutho Haegeman},
	journal={SciPost Phys.},
	volume={10},
	pages={040},
	year={2021},
	publisher={SciPost},
	doi={10.21468/SciPostPhys.10.2.040},
	url={https://scipost.org/10.21468/SciPostPhys.10.2.040},
}

@inbook{Nocedal2006,
title="Calculating Derivatives",
bookTitle="Numerical Optimization",
year="2006",
publisher="Springer New York",
address="New York, NY",
pages="193--219",
isbn="978-0-387-40065-5",
doi="10.1007/978-0-387-40065-5_8",
url="https://doi.org/10.1007/978-0-387-40065-5_8"
}

@article{Mahyaeh2020,
  title = {Study of the phase diagram of the Kitaev-Hubbard chain},
  author = {Mahyaeh, Iman and Ardonne, Eddy},
  journal = {Phys. Rev. B},
  volume = {101},
  issue = {8},
  pages = {085125},
  numpages = {14},
  year = {2020},
  month = {Feb},
  publisher = {American Physical Society},
  doi = {10.1103/PhysRevB.101.085125},
  url = {https://link.aps.org/doi/10.1103/PhysRevB.101.085125}
}

@article{Hassler2012,
doi = {10.1088/1367-2630/14/12/125018},
url = {https://doi.org/10.1088/1367-2630/14/12/125018},
year = {2012},
month = {dec},
publisher = {IOP Publishing},
volume = {14},
number = {12},
pages = {125018},
author = {Hassler, Fabian and Schuricht, Dirk},
title = {Strongly interacting Majorana modes in an array of Josephson junctions},
journal = {New Journal of Physics}
}

@article{Nandy2025,
title = {Quantum dynamics in Krylov space: Methods and applications},
journal = {Physics Reports},
volume = {1125-1128},
pages = {1-82},
year = {2025},
note = {Quantum dynamics in Krylov space: Methods and applications},
issn = {0370-1573},
doi = {https://doi.org/10.1016/j.physrep.2025.05.001},
url = {https://www.sciencedirect.com/science/article/pii/S0370157325001462},
author = {Pratik Nandy and Apollonas S. Matsoukas-Roubeas and Pablo Martínez-Azcona and Anatoly Dymarsky and Adolfo {del Campo}},
}

@book{viswanath1994recursion,
    year = {1994},
    month = {nov},
    publisher = {Springer Berlin Heidelberg},
    author={Viswanath, V.S. and M{\"u}ller, G.},
    series = {Lecture Notes in Physics Monographs},
    title = {The Recursion Method: Application to Many Body Dynamics},
	isbn = {978-3-540-48651-0},
    url = {https://link.springer.com/book/10.1007/978-3-540-48651-0#bibliographic-information},
    edition = {1}
}

@article{Parker2019_B,
  title = {A Universal Operator Growth Hypothesis},
  author = {Parker, Daniel E. and Cao, Xiangyu and Avdoshkin, Alexander and Scaffidi, Thomas and Altman, Ehud},
  journal = {Phys. Rev. X},
  volume = {9},
  issue = {4},
  pages = {041017},
  numpages = {29},
  year = {2019},
  month = {Oct},
  publisher = {American Physical Society},
  doi = {10.1103/PhysRevX.9.041017},
  url = {https://link.aps.org/doi/10.1103/PhysRevX.9.041017}
}

@article{Huse1981,
  title = {Simple three-state model with infinitely many phases},
  author = {Huse, David A.},
  journal = {Phys. Rev. B},
  volume = {24},
  issue = {9},
  pages = {5180--5194},
  numpages = {0},
  year = {1981},
  month = {Nov},
  publisher = {American Physical Society},
  doi = {10.1103/PhysRevB.24.5180},
  url = {https://link.aps.org/doi/10.1103/PhysRevB.24.5180}
}

@article{Huse1983,
title = {Melting and wetting transitions in the three-state chiral clock model},
journal = {Physica A: Statistical Mechanics and its Applications},
volume = {121},
number = {3},
pages = {363-398},
year = {1983},
issn = {0378-4371},
doi = {https://doi.org/10.1016/0378-4371(83)90001-8},
url = {https://www.sciencedirect.com/science/article/pii/0378437183900018},
author = {David A. Huse and Anthony M. Szpilka and Michael E. Fisher}
}

@article{Howes1983,
title = {Quantum model for commensurate-incommensurate transitions},
journal = {Nuclear Physics B},
volume = {215},
number = {2},
pages = {169-208},
year = {1983},
issn = {0550-3213},
doi = {https://doi.org/10.1016/0550-3213(83)90212-2},
url = {https://www.sciencedirect.com/science/article/pii/0550321383902122},
author = {Steven Howes and Leo P. Kadanoff and Marcel {Den Nijs}}
}

@article{Fradkin1980,
title = {Disorder variables and para-fermions in two-dimensional statistical mechanics},
journal = {Nuclear Physics B},
volume = {170},
number = {1},
pages = {1-15},
year = {1980},
issn = {0550-3213},
doi = {https://doi.org/10.1016/0550-3213(80)90472-1},
url = {https://www.sciencedirect.com/science/article/pii/0550321380904721},
author = {Eduardo Fradkin and Leo P. Kadanoff}
}

@article{Auyang1987,
title = {Commuting transfer matrices in the chiral Potts models: Solutions of star-triangle equations with genus>1},
journal = {Physics Letters A},
volume = {123},
number = {5},
pages = {219-223},
year = {1987},
issn = {0375-9601},
doi = {https://doi.org/10.1016/0375-9601(87)90065-X},
url = {https://www.sciencedirect.com/science/article/pii/037596018790065X},
author = {Helen Au-Yang and Barry M. McCoy and Jacques H.H. Perk and Shuang Tang and Mu-Lin Yan}
}

@article{Vongehlen1985,
title = {Zn-symmetric quantum chains with an infinite set of conserved charges and Zn zero modes},
journal = {Nuclear Physics B},
volume = {257},
pages = {351-370},
year = {1985},
issn = {0550-3213},
doi = {https://doi.org/10.1016/0550-3213(85)90350-5},
url = {https://www.sciencedirect.com/science/article/pii/0550321385903505},
author = {G. {von Gehlen} and V. Rittenberg}
}

@article{Fishmann2022,
	title={{The ITensor Software Library for Tensor Network Calculations}},
	author={Matthew Fishman and Steven R. White and E. Miles Stoudenmire},
	journal={SciPost Phys. Codebases},
	pages={4},
	year={2022},
	publisher={SciPost},
	doi={10.21468/SciPostPhysCodeb.4},
	url={https://scipost.org/10.21468/SciPostPhysCodeb.4},
}

@article{JURECA2021,
  title = {{JURECA: Data Centric and Booster Modules implementing the Modular Supercomputing Architecture at JSC}},
  author = {{J{\"u}lich Supercomputing Centre}},
  journal = {Journal of large-scale research facilities},
  volume = {7},
  pages = {A182},
  year = {2021},
  doi = {10.17815/jlsrf-7-182},
  url = {http://dx.doi.org/10.17815/jlsrf-7-182}
}

@misc{Fendley2025,
      title={XYZ integrability the easy way}, 
      author={Paul Fendley and Sascha Gehrmann and Eric Vernier and Frank Verstraete},
      year={2025},
      eprint={2511.04674},
      archivePrefix={arXiv},
      primaryClass={cond-mat.stat-mech},
      url={https://arxiv.org/abs/2511.04674}, 
}

@article{Pancotti2018,
  title = {Almost conserved operators in nearly many-body localized systems},
  author = {Pancotti, Nicola and Knap, Michael and Huse, David A. and Cirac, J. Ignacio and Ba{\~n}uls, Mari Carmen},
  journal = {Phys. Rev. B},
  volume = {97},
  issue = {9},
  pages = {094206},
  numpages = {9},
  year = {2018},
  month = {Mar},
  publisher = {American Physical Society},
  doi = {10.1103/PhysRevB.97.094206},
  url = {https://link.aps.org/doi/10.1103/PhysRevB.97.094206}
}

@dataset{Tausendpfund2026_dataset,
  author       = {Tausendpfund, Niklas and
                  Rizzi, Matteo},
  title        = {Data and Code associated to the paper "Variational
                   Minimization of Almost Strong Zero Modes in
                   Parafermionic Wires"
                  },
  month        = oct,
  year         = 2026,
  publisher    = {Zenodo},
  doi          = {10.5281/zenodo.23281608},
  url          = {https://doi.org/10.5281/zenodo.23281608},
}

@article{Mortier2025_fermionicTN,
	title = {Fermionic tensor network methods},
	pages = {012},
	author = {Mortier, Quinten and Devos, Lukas and Burgelman, Lander and Vanhecke, Bram and Bultinck, Nick and Verstraete, Frank and Haegeman, Jutho and Vanderstraeten, Laurens},
	journal = {SciPost Phys.},
	volume = {18},
	year = {2025},
	publisher = {SciPost},
	doi = {10.21468/SciPostPhys.18.1.012},
	url = {https://scipost.org/10.21468/SciPostPhys.18.1.012}
}
